\documentclass[aps,prb,reprint,superscriptaddress,showpacs]{revtex4-1} 
\usepackage{graphicx,subfigure}
\usepackage{dcolumn}
\usepackage{bm}
\usepackage{amsmath}
\usepackage{amsfonts}
\usepackage{dsfont}
\usepackage{epstopdf}
\usepackage{color}
\usepackage{verbatim}
\newcommand{\bra}[1]{\left\langle #1 \right|}
\newcommand{\ket}[1]{\left|#1\right\rangle}

\begin{document}

%
\title{Demonstration of Open Quantum System Optimal Control in Dynamic Nuclear Polarization}

%
\author{S. Sheldon} 
\affiliation{Department of Nuclear Science and Engineering, MIT, Cambridge, MA 02139, USA}
\affiliation{Institute for Quantum Computing, University of Waterloo,
Waterloo, Ontario N2L 3G1, Canada}
\affiliation{IBM T.J. Watson Research Center, Yorktown Heights, NY 10598, USA}
\author{D.G. Cory}
\affiliation{Institute for Quantum Computing, University of Waterloo,
Waterloo, Ontario N2L 3G1, Canada}
\affiliation{Canadian Institute for Advanced Research, Toronto,
Ontario M5G 1Z8, Canada}
\affiliation{Department of Chemistry, University of Waterloo,
Waterloo, Ontario N2L 3G1, Canada}
\affiliation{Perimeter Institute for Theoretical Physics, Waterloo,
Ontario N2L 2Y5, Canada}

\date{\today}

\begin{abstract}

Dynamic nuclear polarization (DNP) is used in nuclear magnetic resonance (NMR) to transfer polarization from electron spins to nuclear spins.  The resulting nuclear polarization enhancement can, in theory, be two or three orders of magnitude depending on the sample.  In solid state systems, however, there are competing mechanisms of DNP, which, when occurring simultaneously, reduce the net polarization enhancement of the nuclear spin.  We present a simple quantum description of DNP and apply optimal control theory (OCT) with an open quantum system framework to design pulses that select one DNP process and suppress the others.  We demonstrate experimentally an order of magnitude improvement in the DNP enhancement using OCT pulses. 

\end{abstract}

%
\maketitle
\section{introduction and motivation}

Dynamic nuclear polarization (DNP) has long been used in NMR to increase the signal to noise of an experiment by 
transferring polarization from 
electron spins to nuclear spins \cite{abragam78}.  The thermal polarization of an NMR sample is proportional to the external magnetic field, $\vec{B}_0$, the 
gyromagnetic ratio, $\gamma$, of the spins in the sample, and is inversely proportional to the temperature, $T$, in the high temperature limit, 
$P \approx \gamma \vec{B}_0/k_BT$.  
The electron gyromagnetic ratio, $\gamma_e$, is two to three orders of magnitude larger than $\gamma_n$, the gyromagnetic ratio for a given nuclear spin. 
Thus electron spins are more highly polarized at thermal equilibrium than nuclear spins.
DNP  has become a valuable tool for NMR spectroscopy, medical
imaging, magnetic sensors, and semiconductor spin studies due to the increased sensitivity provided by nuclear polarization enhancements
\cite{dementyev08, dementyev11, king10}.  

While DNP has been studied extensively \cite{abragam78, overhauser53, carver56}, a full quantum description has been missing
from the literature.  In fact, DNP offers a illustrative example of open quantum system dynamics. Here we consider a two-spin electron-nuclear hyperfine coupled spin system, in which the 
polarization transfer can occur through several pathways.  Some of these DNP processes require decoherent mechanisms, so it is useful to
consider the pathways of DNP to be various information pathways in a more general open quantum framework. Often one may want
to select a particular pathway without also addressing unwanted transitions. 
When the spin transitions are close or there is an overlap between the desired excitation frequencies and other transitions in the system, sophisticated
control techniques are required. 

In this work we applied optimal control theory (OCT) to find
control sequences that select one DNP pathway to produce the largest possible nuclear polarization enhancement.
OCT is a valuable tool in magnetic resonance and for control of quantum information \cite{khaneja01, fortunato02, khaneja05}. Optimal control pulses have
been used previously to demonstrate universal
control in electron-nuclear systems with microwave irradiation only\cite{hodges08}. 

We use DNP to demonstrate the improved control available  in open quantum systems through optimal control methods. 
For unitary systems, there is always a perfect solution with  optimal control, even in systems with a distribution of Hamiltonians such as powder samples or samples in the presence of large field inhomogeneities. Finding optimal solutions becomes more complex when decoherent processes are involved.  
There has been theoretical work applying OCT to open quantum systems \cite{glaser,rabitz,koch2014}, including optimization of DNP processes 
with simultaneous RF and microwave control \cite{maximov08,li09}.  Others have also studied optimal control in the context of quantum measurement \cite{egger2014}, quantum systems coupled to non-Markovian environments \cite{wilhelm2009}, and minimizing noise \cite{kosloff, whaley}.  Nonunitary control of DNP has also been considered in quantum dots \cite{barnes,economou}.
We present here an experimentally realized a scheme for microwave only optimal control of DNP in
an electron-nuclear system with relaxation mechanisms included. 

\begin{figure}[!bt]
	\centering
		\includegraphics[width=8.5cm]{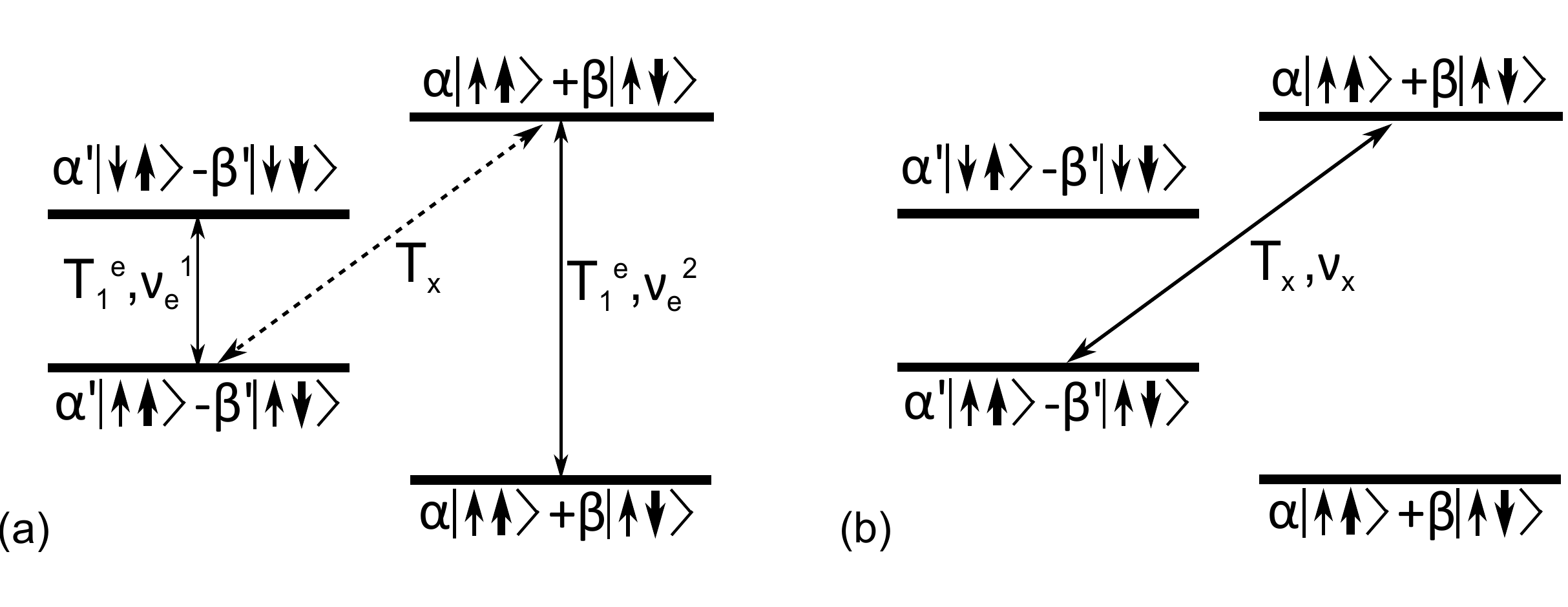}
		\caption{The energy level diagram on the left illustrates the Overhauser effect: the two electron resonances, $\nu_e^1$ and $\nu_e^2$, are saturated, and
		polarization is transferred through the zero quantum cross relaxation, $T_x$. The diagram on the right represents the solid effect:
		the zero quantum transition, $\nu_x$, is directly irradiated.}
	\label{fig:EnergyLevels}
\end{figure}

Our system consists of one electron hyperfine coupled to one hydrogen nucleus, forming a four level system (Fig. \ref{fig:EnergyLevels}).  There are two available DNP pathways, depending on which transitions are
excited. Applying microwave irradiation on resonance with the zero quantum transition induces an electron-nuclear mutual spin flip-flop, which directly transfers
polarization from the electron spin to the nuclear spin \cite{abragam78}.  This is known as the solid effect (SE).  Alternatively, the Overhauser effect (OE) indirectly causes this flip-flop through a cross relaxation process that occurs when both electron resonances are saturated (see Fig. \ref{fig:EnergyLevels} for diagrams of both mechanisms) \cite{overhauser53}.  The zero quantum transition is ordinarily forbidden, and cross relaxation ($T_x$ in Fig. \ref{fig:EnergyLevels}) is not allowed.  Both the solid effect and Overhauser effect can occur, however, when the nuclear Zeeman states are mixed due to an anisotropic hyperfine coupling.  

Under continuous wave (CW) irradiation all transitions are excited to varying degrees, driving both processes. The OE leads to a positive nuclear polarization, and the SE to negative nuclear polarization.  Consequently if both the OE and SE are driven, the net polarization enhancement is less than the polarization achievable by 
either process acting on the system alone.  
It is therefore advantageous to find a control sequence that selects one DNP process
and suppresses the other.  A pulse sequence capable of driving exclusively one of these two opposing methods of DNP must account
for both decoherent processes and unitary evolution.  

The theoretical nuclear polarization enhancement is 
limited by the ratio $\gamma_e/\gamma_n$ \cite{jeffries}.  For hydrogen, the maximum achievable enhancement is 660. 
In any physical system the enhancement will be less than the theoretical limit due to a factors 
such as imperfect saturation, leakage effects from relaxation, and 
the competition between various DNP processes.

\section{Quantum Description of DNP}
Historically DNP has been described using a semi-classical model with systems of rate equations.  We have the tools, however, to analyze the polarization transfer
as a fully quantum process with system-environment interactions.  We can write both the solid effect 
and Overhauser effect as quantum maps, and we do so here using a Kraus operator representation.  
The minimum set of operators necessary to determine these maps are the drift Hamiltonian, $H_{drift}$, the control Hamiltonian,
$H_{control}$, a set of Kraus operators for the $T_1^e$ relaxation of the electron, and Kraus operators for $T_x$, the zero quantum cross relaxation between
the electron and the nucleus.  

The drift Hamiltonian is the same for both DNP mechanisms.  $H_{drift}$ is the sum of the Zeeman interactions for the electron and the nucleus and the hyperfine
coupling between the two spins. We assign a particular Hamiltonian, which contains a hyperfine coupling with both isotropic ($A$) and anisotropic ($B$) terms.   In the high field limit the electron Zeeman term will be much larger than the other terms, we take the secular approximation with respect to the electron spin quantum number
and the drift Hamiltonian is
\begin{equation}
H_{drift} = \omega_S S_z\ +\omega_I I_z + A S_zI_z + B S_zI_x
\label{eq:driftHamiltonian}
\end{equation}
where $\vec{S}$ and $\vec{I}$ are the spin operators for the electron and the nucleus respectively,  
$\omega_S$/$\omega_I$ the electron/nucleus Larmor frequency. 
For efficient polarization transfer, it is necessary to use a system in which the anisotropic hyperfine is on the same order as the nuclear Zeeman term,
 so that the eigenstates are mixtures of nuclear spin up and spin down and the zero quantum transition is weakly allowed \cite{abragam78}.

The control Hamiltonian for the OE is produced by microwave
irradiation on resonance with the electron spin flip transitions. In the rotating frame of the electron this is:
\begin{equation}
H_{OE} = \omega_d S_x\otimes\mathds{1}  
\label{eq:OEcontrolHamiltonian}
\end{equation}
where $\omega_d$ is the Rabi frequency of the drive field. Similarly for the SE, the microwave control on resonance with the zero quantum transition yields the control Hamiltonian:
\begin{equation}
H_{SE} = {\omega_d}\left(S_xI_x +S_yI_y\right)
\label{eq:SEcontrolHamiltonian}
\end{equation}

The remaining operators needed to mathematically describe DNP are the non-unitary operators that drive the decoherent processes.  The SE requires only 
$T_1^e$ relaxation, while both $T_1^e$ and $T_x$ are necessary for the OE. All the unitary and non-unitary operators for the solid and Overhauser effects are listed in Table \ref{tab:MapsForBothMechanisms}.

To illustrate that these operators are sufficient, we consider the following discrete maps which increase the nuclear polarization. Here we assume that $T_1$ of the nuclear spin is much longer than $T_1^e$ and can be neglected.  

For the Overhauser effect:
\begin{equation}
\Lambda_{OE} = [U_{OE}] \rightarrow [T_x] \rightarrow [T_1^e]
\end{equation}
\begin{enumerate}
\item $U_{OE}$ saturates the electron spins, removing any electron magnetization.
\item $T_x$ drives population to the nuclear spin $\left|\downarrow\right>$ state.
\item $T_1^e$ removes entropy from the system.
\end{enumerate}

And the Solid effect:
\begin{equation}
\Lambda_{SE} = [U_{SE}] \rightarrow [T_1^e]
\end{equation}
\begin{enumerate}
\item $U_{SE}$ transfers polarization from the electron through the zero quantum transition.
\item $T_1^e$ removes entropy from the system and resets the electron spin.
\end{enumerate}

\begin{table}
	\centering
		\begin{tabular}{ccc}\hline\hline
		& Overhauser Effect & Solid Effect \\ \hline
		& &\\
		\vspace{5mm}
		$U_{drift} $& $e^{(\omega_I I_z + A S_zI_z + B S_zI_x )}$& $e^{(\omega_I I_z + A S_zI_z + B S_zI_x )}$\\
		\vspace{5mm}
		$U_{control}$ & $e^{\left(-i\frac{\pi}{4}\omega_d S_x\otimes\mathds{1}  \right)}$&$ e^{\left(-i\frac{\pi}{4}\frac{\omega_d}{2}\left(S_xI_x +S_yI_y\right)\right)}$\\
		\vspace{5mm}
		Relaxation& $T_1^e$, $T_x$ & $T_1^e$ \\		
		\hline\hline
		\end{tabular}
	\caption{Maps for both mechanisms in rotating frame of the electron.}
	\label{tab:MapsForBothMechanisms}
\end{table}

In a physical system decoherence is continuous, and we cannot sequentially apply the unitary processes then decoherent processes compose these discrete maps.  Instead we can move to a continuous
time map and find the generators for DNP.  We choose the Kraus operator formalism to describe the decoherent processes and determine the continuous map. 

In the Kraus operator form, a set of operators $\{M_k\}$ act on the state $\rho$ yielding the map,

\begin{equation}
\rho' = \sum_k M_k\rho M_k^{\dagger}
\label{eq:KrausMap1}
\end{equation} 
This map is completely positive and trace-preserving if the Kraus operators satisfy the condition,
\begin{equation}
\sum_kM_k^{\dagger}M_k = \mathbb{I}
\label{eq:krausCondition}
\end{equation}

We start with the Kraus operators that describe $T_1$ relaxation.  During the $T_1$ process, the electron spin flips while the nuclear spin remains in the same state (indicated by $T_1^e$ Fig. \ref{fig:EnergyLevels}).  
 A set of Kraus operators driving 
this type of relaxation must act as the operator $S_x \otimes \mathds{1}_I$ in 
the energy eigenbasis.   We also require the Kraus operators for a $T_1$ process to
return the system (or relevant subsystem) to thermal equilibrium.  The following set of  Kraus operators describes $T_1$ relaxation on the electron
spin:

\begin{align}
&A_1 = \sqrt{p}\left(\begin{array}{cc}1&0\\0&\sqrt{\epsilon}\end{array}\right)\otimes \mathbb{I} \nonumber\\
& A_2 = \sqrt{p}\left(\begin{array}{cc}0&\sqrt{1-\epsilon}\\0&0\end{array}\right)\otimes \mathbb{I}\nonumber\displaybreak[0]\\
&A_3 = \sqrt{1-p}\left(\begin{array}{cc}\sqrt{\epsilon}&0\\0&1\end{array}\right)\otimes \mathbb{I} \nonumber\\
& A_4 = \sqrt{1 - p}\left(\begin{array}{cc}0&0\\\sqrt{1-\epsilon}&0\end{array}\right)\otimes \mathbb{I}\nonumber\\
\label{eq:T1operators}
\end{align}
where the parameter $\epsilon = \exp(-t/T_1^e)$ sets the rate of the process and $p = \exp(-h\omega_S /2k_BT)$ determines the final polarization \cite{nandc}.
 
The other set of Kraus operators that is required to describe DNP is that which implements the electron-nuclear cross relaxation, $T_x$ (see Fig. \ref{fig:EnergyLevels}).  
This is essentially a $T_1$ process that acts on the zero quantum subspace, a depolarizing channel
that leads to transitions 
$\ket{\downarrow\beta}\Leftrightarrow\ket{\uparrow\alpha^{\perp}}$.  As in the electron $T_1$ case the system should return to thermal equilibrium.  
The following operators satisfy these requirements:
\begin{align}
&B_1 = \sqrt{p_{zq}}\left(\begin{array}{cccc} 
1&0&0&0\\
0&1&0&0\\
0&0&\sqrt{\epsilon_{zq}}&0\\
0&0&0&1\\
\end{array}\right)\nonumber\\
& B_2 = \sqrt{p_{zq}}\left(\begin{array}{cccc}
0&0&0&0\\
0&0&\sqrt{1-\epsilon_{zq}}&0\\
0&0&0&0\\
0&0&0&0\\
\end{array}\right)
\nonumber\displaybreak[0]\\
&B_3 = \sqrt{1 - p_{zq}}\left(\begin{array}{cccc}
1&0&0&0\\
0&\sqrt{\epsilon_{zq}}&0&0\\
0&0&1&0\\
0&0&0&1\\
\end{array}\right)\nonumber\\
& B_4 = \sqrt{1 - p_{zq}}\left(\begin{array}{cccc}
0&0&0&0\\
0&0&0&0\\
0&0&\sqrt{1-\epsilon_{zq}}&0\\
0&0&0&0\\
\end{array}\right)
\nonumber\\
\end{align}
with $p_{zq} = \exp(-h(\omega_S-\omega_I)/2k_BT)$ and $\epsilon_{zq} = \exp(-t/T_{ZQ})$, where $T_{ZQ}$ is the zero quantum cross relaxation.
These Kraus operators satisfy the condition $\sum_kB_k^{\dagger}B_k = 1$.


We now have the Kraus map for the total evolution
\begin{equation}
\rho' = \sum_kA_k\left[\sum_lB_le^{-iHt}\rho e^{iHt}B_l^{\dagger}\right]A_k^{\dagger}
\end{equation}
which we rewrite in terms of a single set of Kraus operators, $\{M_k\}$,
\begin{equation}
\rho' = \sum_k M_k\rho M_k^{\dagger}
\end{equation} 



The next step is to find the reduced map that acts on the nuclear spin subsystem and to write it in the Kraus form.  
We start by transforming the Kraus map to a supermatrix representation that acts on the vectorized density matrix, $\hat{\hat{\rho}}$,
\begin{equation}
\hat{\hat{\rho}}' = \textbf{S}\hat{\hat{\rho}}
\label{eq:supermatrix}
\end{equation}
In a general sense, we can consider the electron to be the environment of the nucleus.  We evaluate the reduced map on the nuclear spin by tracing over this 
``environment''. This is accomplished by applying $\textbf{S}$ to a known state of the environment and taking the partial trace,
\begin{equation}
\textbf{S}_n = Tr_E(\textbf{S}\hat{\hat{\rho}}_E\otimes\hat{\hat{\rho}}_n)
\label{eq:traceEnvironment}
\end{equation}

The Kraus operators acting on the reduced space can be found from the Choi matrix, which is given by a reshuffling of the reduced supermatrix, $\textbf{S}_n$ \cite{wood}.
The Choi matrix is defined as 
\begin{equation}
\Lambda_C = (\mathbb{I}\otimes\Lambda)\sum_{ij}E_{ij}\otimes E_{ij} = \sum_{ij}E_{ij}\otimes\Lambda(E_{ij})
\label{eq:choiMatrix}
\end{equation}
where $E_{ij} = \ket{i}\bra{j}$ and the vectors $\{\ket{i},\ket{j}\}$ form a basis of the Hilbert space. 
The procedure for extracting Kraus operators from the Choi matrix follows. We rewrite $\Lambda_C$ in terms of its normalized eigenvectors, $\ket{a_k}$,
\begin{equation}
\Lambda_C = \sum_k\ket{a_k}\bra{a_k}
\end{equation}
The columns of $\Lambda_C$ form a set of columnized Kraus operators $\{A_k\}$  satisfying the condition of Eq. (\ref{eq:krausCondition})
with $d^2$ operators in the set, where $d$ is the dimension of the Hilbert space.  

Using this procedure we find the following Kraus operators for the nuclear spin map.  The final Kraus operators for both DNP processes are similar in form
to the $T_1$ Kraus operators of Eq. (\ref{eq:T1operators}) and do in fact describe polarizing channels on the nuclear spin.  For the solid effect 
effect the four Kraus operators found have the structure:
\begin{align}
&M_1 = \alpha\left(\begin{array}{cc}
0&\sqrt{\Gamma}\\
0&0\\
\end{array}\right)\nonumber\\
&M_2 = \alpha\left(\begin{array}{cc}
0&0\\
\sqrt{1-\Gamma}&0\\
\end{array}\right)\nonumber\\
&M_3 = \beta_{-}\left(\begin{array}{cc}
\Delta_{-}&0\\
0&1\\
\end{array}\right)\label{eq:SEkraus}	\\ 
&M_4 = \beta_{+}\left(\begin{array}{cc}
\Delta_{+}&0\\
0&1\\
\end{array}\right)\nonumber\\\nonumber
\end{align}
\vspace{-10pt}
where the parameters $\Gamma$, $\alpha$,  $\beta_{\pm}$, and $\Delta_{\pm}$ depend on the Hamiltonian parameters and $T_1$ as follows,
\begin{gather}
\Gamma = \gamma_1=\exp{\frac{\hbar\omega_S}{2k_BT}}\label{eq:SEparameters}\\
\alpha = \sqrt{\frac{(1-e^{-t/T_1})}{2}}\nonumber\\
\beta_{\pm} = \frac{(\mp 4-3\sqrt{2})(t-4T_1)}{8\sqrt{3\pm 2\sqrt{2}} T_1} +O(t^2)\nonumber\\
\begin{align*}
\Delta_{\pm}  &= \frac{-8T_1+t\left(2\pm \sqrt{2} +  i 4 T_1 \left(\eta_{-}+\eta_{+}\right)\mp 2\sqrt{2}\gamma_1 \right)}{2(t-4T_1)}\nonumber\\
&\hspace{6cm}+O(t^2)\nonumber
\end{align*}\nonumber
\end{gather}
with 
\begin{align}
\eta_{\pm} &= \sqrt{4A^2+4B^2 \pm 4A\omega_I+\omega_I^2}
\label{eq:eta}
\end{align}
$\beta_{\pm}$ and $\Delta_{\pm}$ are approximated to first order in time, $t$.  This approximation is valid under the assumption that the time step is much shorter than the relaxation time, $T_1$.
 
Similarly for the Overhauser Effect, we find the reduced Kraus operators,
\begin{align}
&M'_1 = \alpha'\left(\begin{array}{cc}
0&\sqrt{\Gamma_1(1-\Gamma_x)}\\
0&0\\
\end{array}\right)\nonumber\\
&M'_2 = \alpha'\left(\begin{array}{cc}
0&0\\
\sqrt{(1-\Gamma_1)\Gamma_x}&0\\ 
\end{array}\right)\nonumber\\
&M'_3 = \beta'_{-}\left(\begin{array}{cc}
\Delta'_{-}&0\\
0&1\\
\end{array}\right)\\
&M'_4 = \beta'_{+}\left(\begin{array}{cc}
\Delta'_{+}&0\\
0&1\\
\end{array}\right)\nonumber\\\nonumber
\end{align}
with the parameters,
\begin{gather}
\Gamma_1 = 1 -2\gamma_1(1-e^{t/T_1})\nonumber\\
\Gamma_x = \gamma_x =\exp{\frac{\hbar(\omega_S-\omega_I)}{2k_BT}} \\
\alpha' = e^{-t\left(1/T_1+1/T_x\right)}\left(e^{t/T_x}-1\right)\nonumber\\ \nonumber
\end{gather}
\vspace{-5pt}
and
\vspace{0pt}
\begin{widetext}
\begin{gather}
\beta'_{\pm} = \sqrt{\frac{-t\mp 4 T_x\pm \sqrt{4\times(-4tT_x+8T_x^2+t^2(1+2T_x^2(\eta_-+\eta_+)^2-2\gamma_x(1-\gamma_x)))}}{4T_x}} +O(t^2)\nonumber\\
\Delta'_{\pm}  = \frac{2i\left(\pm t(1-2\gamma_x)+\sqrt{4\times(-4tT_x+8T_x^2+t^2(1+2T_x^2(\eta_-+\eta_+)^2-2\gamma_x+2\gamma_x^2))}\right)}{8iT_x-t(2i+4Tx(\eta_-+\eta_+))}+O(t^2)\nonumber
\end{gather}
\end{widetext}
with $\gamma_1$ and $\eta$ as defined in Eqs. (\ref{eq:SEparameters}) and (\ref{eq:eta}) respectively.  Again, for simplicity, we have taken $\beta'_{\pm}$ and $\Delta'_{\pm}$ to first order in $t$ with the assumption that $t$ is shorter than $T_1$ and $T_x$.  If we make the same approximation for $\Gamma_1$ and $\Gamma_x$, then the cross relaxation term dominates and the Kraus operators $M'_1$ and $M'_2$ take the same form as those for the solid effect ($M_1$ and $M_2$ in Eq.~(\ref{eq:SEkraus})) but with the direction of polarization reversed:
\begin{align} 
&M'_1 \rightarrow \sqrt{\frac{t}{2T_x}}\left(\begin{array}{cc}
0&\sqrt{1-\gamma_x}\\
0&0\\
\end{array}\right)\nonumber\\
&M'_2 \rightarrow \sqrt{\frac{t}{2T_x}}\left(\begin{array}{cc}
0&0\\
\sqrt{\gamma_x}&0\\ 
\end{array}\right)\nonumber\\
\end{align}
These Kraus operators satisfy the requirement of Eq. (\ref{eq:krausCondition}) for a CPTP map. 
$\{M_k\}$ and $\{M'_k\}$ resemble the Kraus operators for a return to thermal equilibrium, each with a new equilibrium defined by $\gamma_1$ and $\gamma_x$.  These reduced maps on the nuclear spin drive the spin to a hyperpolarized state, and with opposite sign for the solid effect versus the Overhauser effect.

\begin{figure}
	\centering
		\includegraphics[scale=0.5]{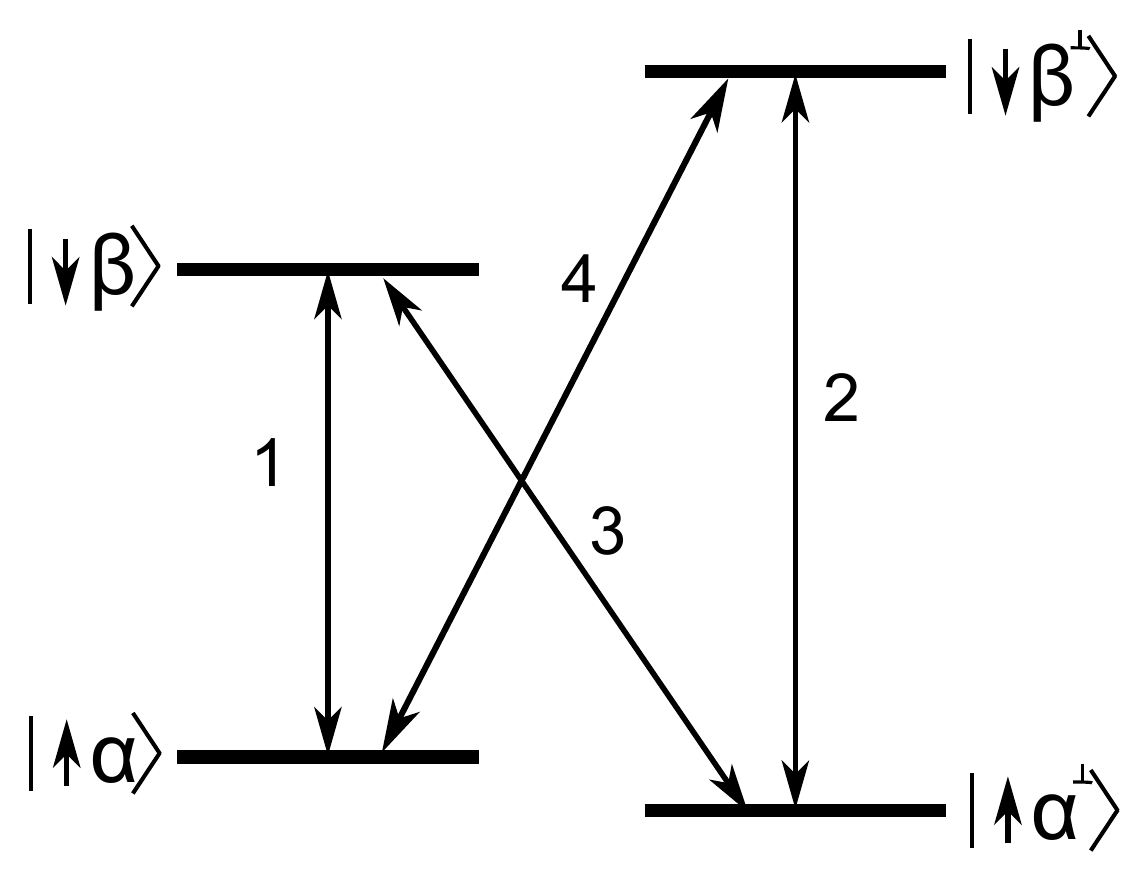}
	\caption{The four possible transitions in the two spin system.  1 and 2 are electron only transitions.  3 is the zero quantum transition,
	and 4 the double quantum.}
	\label{fig:energyLevelDiagramTransitionLabels}
\end{figure}

\section{Optimal Control Theory}\label{section:OCT}
The generators given above assume perfect control.  In reality, we cannot produce the perfect unitaries $U_{SE}$ and $U_{OE}$.  Any microwave field
that we apply to the system will excite all four transitions to some degree, resulting in non-ideal unitaries that will drive the Overhauser effect and solid effect
simultaneously and reduce the final nuclear polarization.  In order to increase selectivity, we incorporated optimal control theory (OCT).

OCT allows the design of pulse sequences that perform any desired unitary on the four level system.  There are many unitary operators that result in a hyperpolarized
nuclear state. Exciting the transitions that drive the Overhauser effect, for example, requires a unitary that acts only on the subspace of the electron spin, whereas the solid effect is
driven by unitary operators acting on the zero quantum subspace.  For the experimental implementation we choose a single operator to optimize towards; in principle
we can optimize to any operator on the electron subspace that leads to a polarization transfer.

\begin{figure}
	\centering
		\includegraphics[width=8cm]{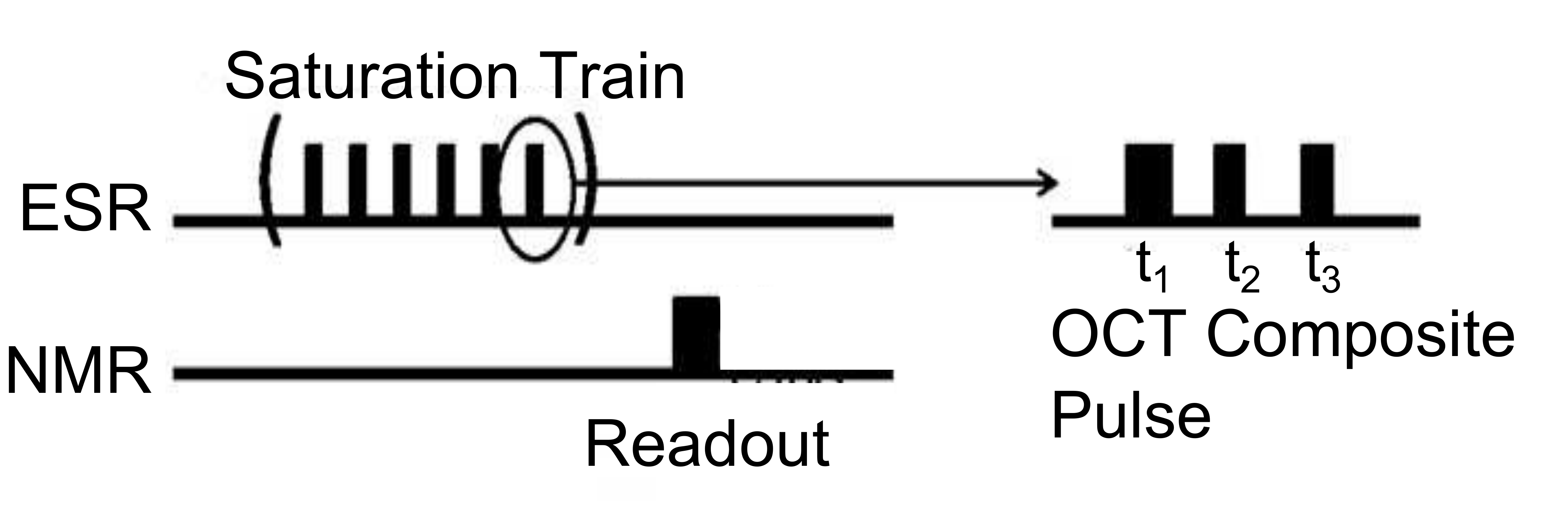}
	\caption{Timing diagram of the experiment: the $\pi/2$ saturation is applied for a variable length of time followed by the NMR readout pulse.}
	\label{fig:OCTDNPexperiment}
\end{figure}

There are four possible transitions in the hyperfine coupled electron-nuclear system, as shown in Fig. \ref{fig:energyLevelDiagramTransitionLabels}.
Depending on which transitions are driven during the DNP experiment, the effective unitary will lead to a combination of the OE and SE.  
As will be explained in the next section, the pulse sequence used to drive DNP in this work is a saturation train (Fig. \ref{fig:OCTDNPexperiment}), where each pulse will either be a hard pulse or a composite OCT pulse.  The angle of rotation that a single pulse performs on each transition in Fig. \ref{fig:energyLevelDiagramTransitionLabels} determines the degree of saturation of that transition.

\begin{figure*}
\centering
$\begin{array}{cc}
\includegraphics[width=0.4\linewidth]{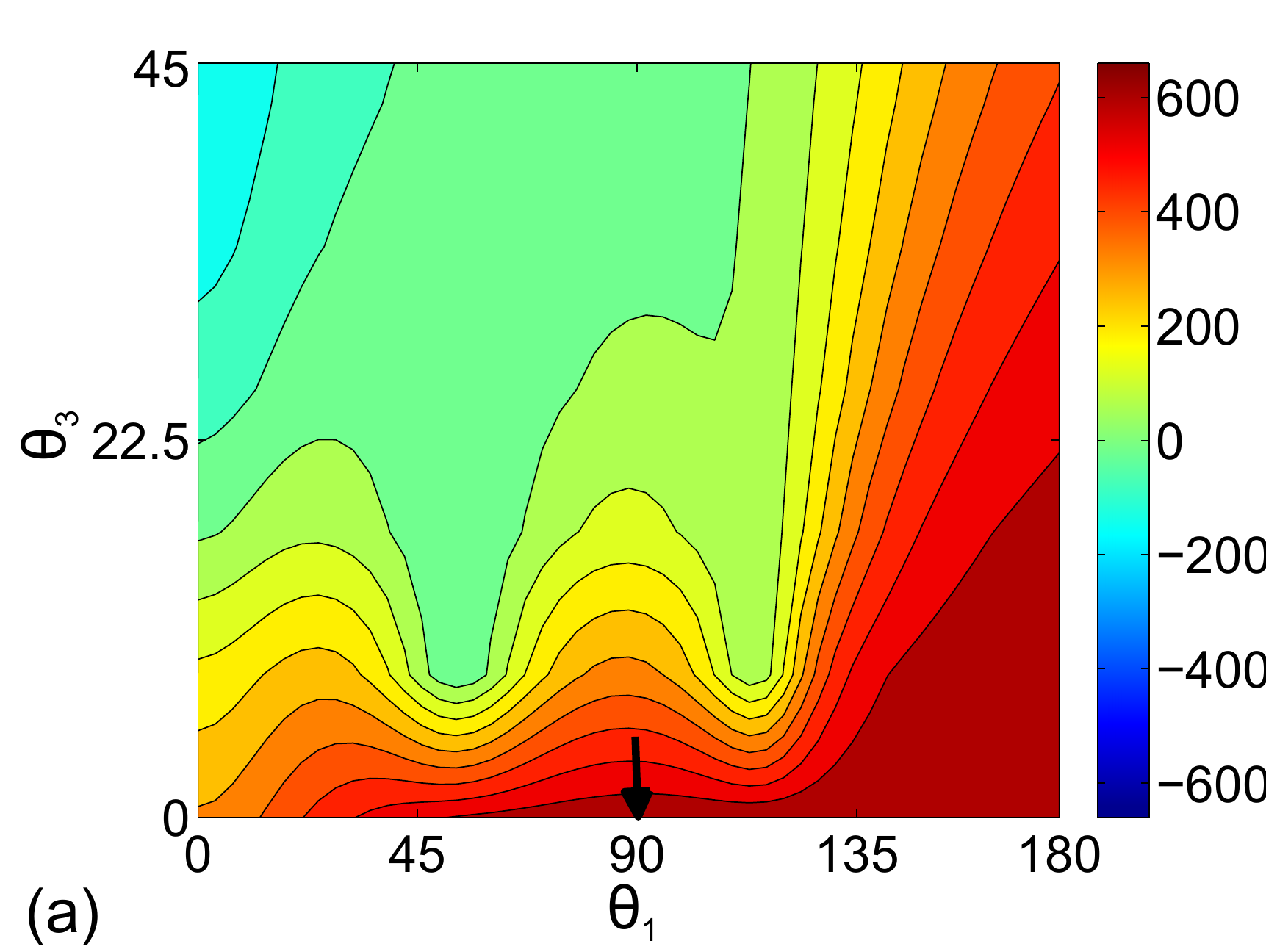}&\hspace{5pt}\includegraphics[width=0.41\linewidth]{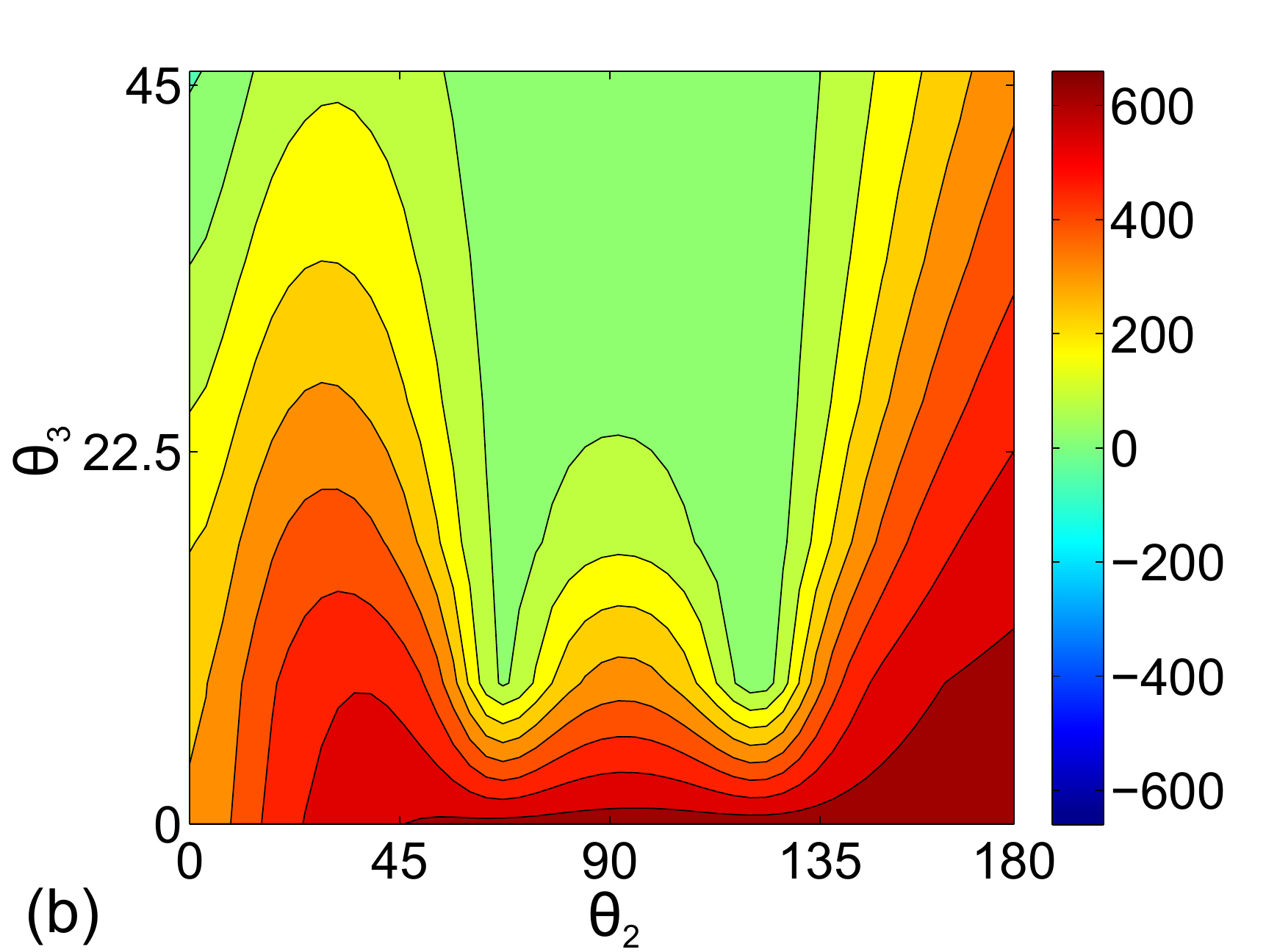}\\
\includegraphics[width=0.4\linewidth]{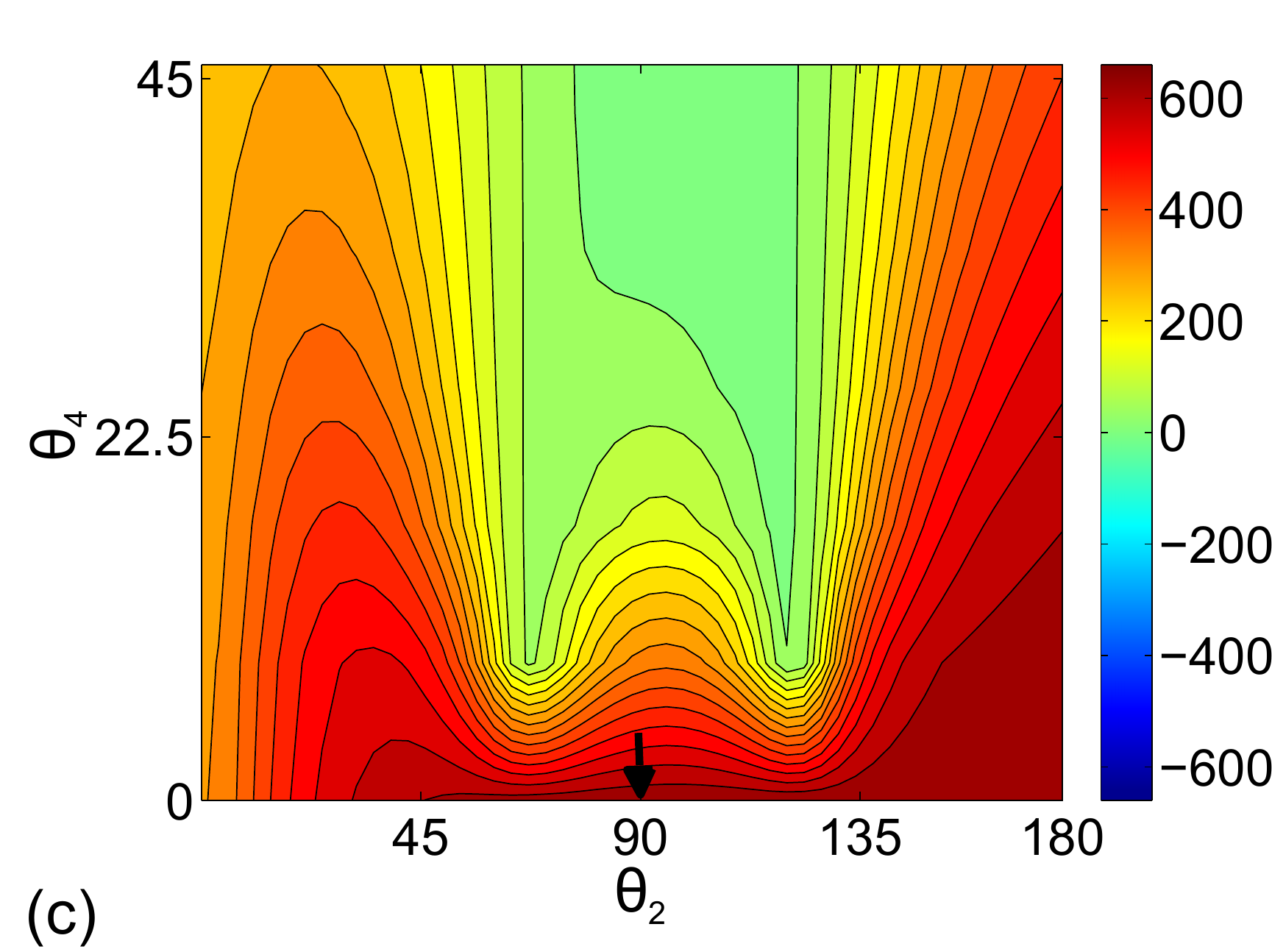} &\includegraphics[width=0.4\linewidth]{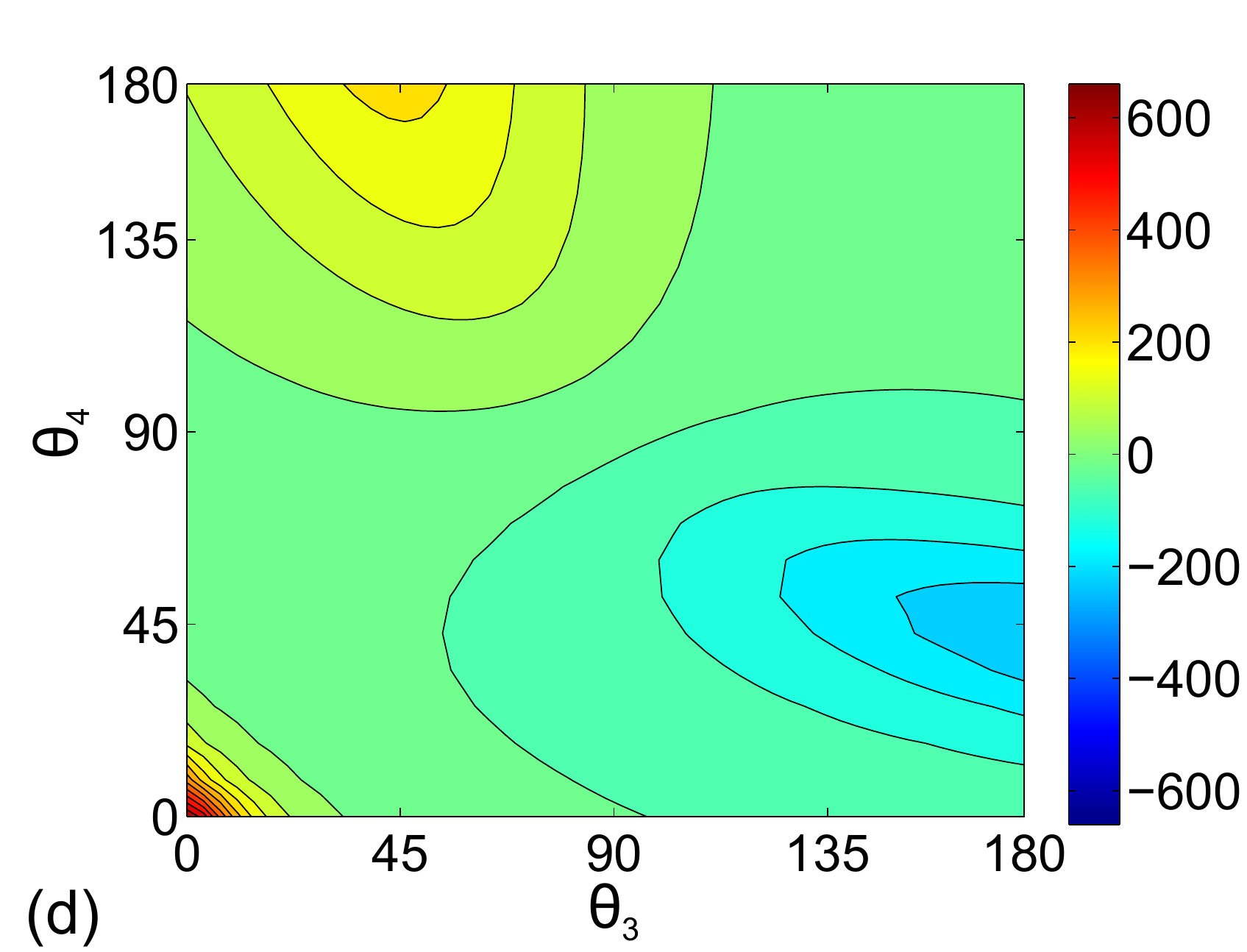}\\
\includegraphics[width=0.4\linewidth]{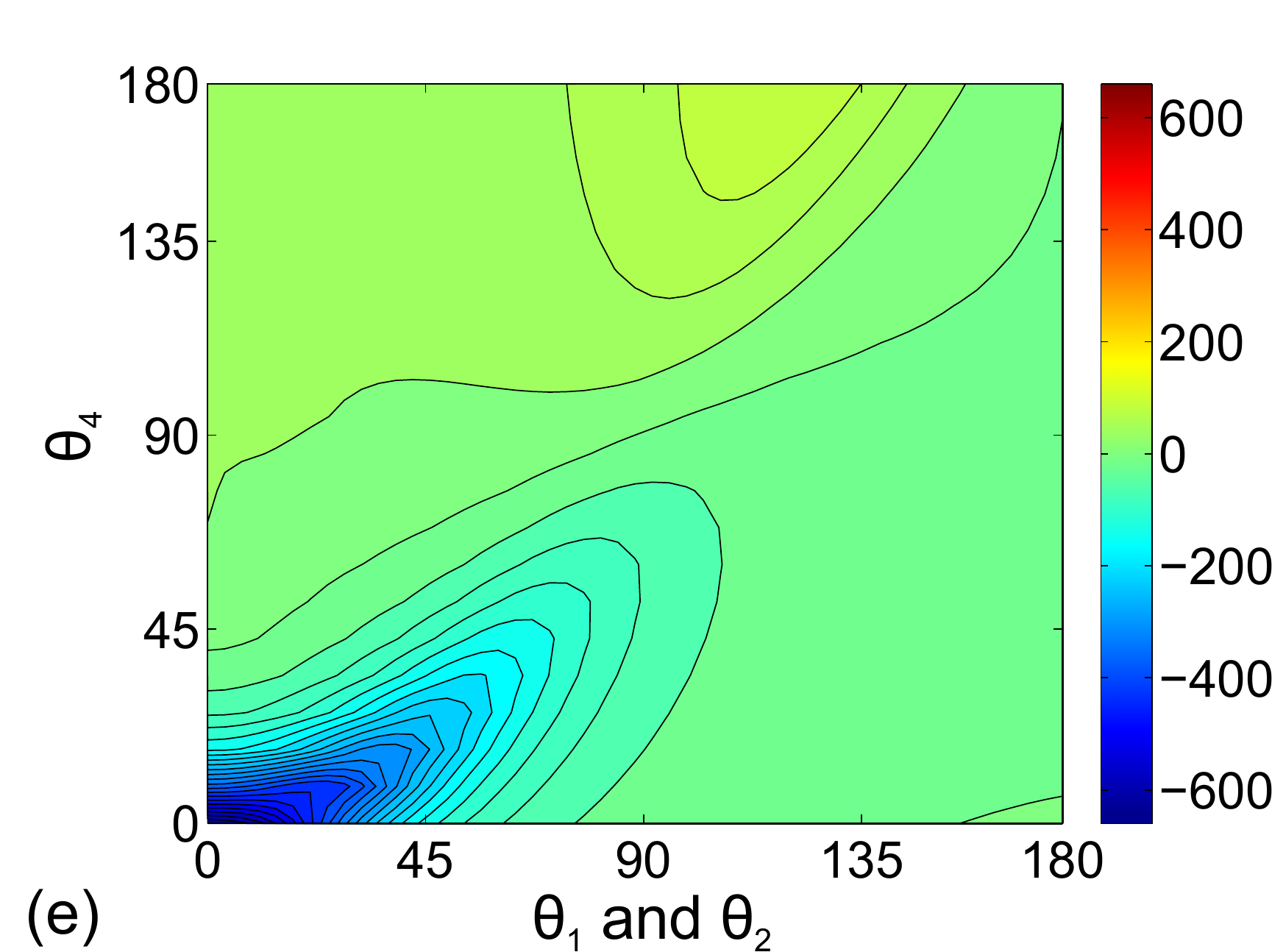}
\end{array}$
\caption{Color map indicates the DNP enhancement, with the maximum of 660 in red and minimum of -660 in blue.  The x and y axes indicate the degree of rotation for a repeated train of pulses (as in Fig. \ref{fig:OCTDNPexperiment})) on the transition given by the axes labels. The angles $\theta_i$ correspond to the transitions in Fig. \ref{fig:energyLevelDiagramTransitionLabels}.  The following cases are depicted: (a) Increasing saturation on transitions 1 and 3 while applying $\pi/2$ pulses to 2; (b) Increasing the 2 and 3, with $\pi/2$ pulses on the other electron resonance, 1; (c) increasing 2 and 4 with full $\pi/2$ saturation on 1; (d) increasing 3 and 4 with both electron resonances saturated; and (e) increasing the electron resonance simultaneously and increasing 4 with $\pi/2$ pulses on 3.  The ideal case for the Overhauser effect is occurs at the points indicated by arrows in (a) and (c).}
\label{fig:DNPspace2}
\end{figure*}

By varying the rotation angle of each transition we can explore the space of DNP and find points in this space determined that yield a nuclear polarization enhancement. Fig. \ref{fig:DNPspace2} presents the DNP enhancement of the nuclear spin signal for varying saturation levels of the four transitions.    There are four angles $\theta_i$ corresponding to the transitions of Fig. \ref{fig:energyLevelDiagramTransitionLabels}.	
These plots illustrate several representative slices of the full space of DNP.  
The x-axes of (a)-(c) indicate the rotation angle of the pulses in Fig. \ref{fig:OCTDNPexperiment} on transitions 1 and 2 (i.e. the transitions that involve flipping only the electron), 
with no excitation of the zero quantum and double quantum transitions.  These points in the space indicate unitaries which drive the Overhauser effect only and lead to positive nuclear polarization enhancement. When the cross transitions (3 and 4 in Fig. \ref{fig:energyLevelDiagramTransitionLabels}) are also excited (increasing saturation corresponds to larger rotations on the y-axes), the microwave control also induces the solid effect and drives the 
nuclear spin towards a negative polarization. In regions where single quantum transitions and cross transitions are both excited, the net nuclear polarization is less than the theoretical maximum of 660.

Experimentally it is difficult to drive polarization enhancement through the solid effect.  The zero quantum and double quantum 
transitions are only weakly allowed, and therefore require large microwave powers to saturate.   
In this work we choose instead to drive the Overhauser effect by optimizing $\pi/2$ rotations on the $1$ and $2$ transitions with no excitation on the cross transitions, $3$ and $4$. This case is indicated by arrows on figures (a) and (c) of Fig. \ref{fig:DNPspace2}.  



\section{pulse finding}

We accomplish the saturation of the electron resonances for the OE through a $\pi/2$ microwave pulse train as shown in Figure \ref{fig:OCTDNPexperiment}.  
The goal of the pulse finder is to produce composite pulses that perform the $\pi/2$ rotation on both electron resonances without any rotation
on the zero quantum (or double quantum) subspace. The desired unitary for Overhauser DNP is  
\begin{equation}
U = \exp(-i \pi/2  S_x \otimes \mathds{1})
\label{eq:unitaryOE}
\end{equation}

Our pulse optimization method uses the Nelder-Mead simplex algorithm to numerically search over
the control parameter space \cite{nelder65}.  The pulses are restricted to on/off modulation; 
the total pulse consists of several time steps during which the microwave control is either 
on or off and the lengths of each step are left as parameters. 

To optimize the pulses, we first find the map that the OCT pulse performs when relaxation is present in the system
 and compare this to the map produced by the desired $\pi/2$ rotation.
We have a set of operators
$\{M_k\}$ such as in Eq. (\ref{eq:KrausMap1}) for each time step.  
The successive application of these Kraus operators for each segment of the pulse gives the map for the composite pulse:
\begin{equation}
\rho' = \sum_{n_1}M^1_{n_1}\left[\cdots\sum_{n_N}M^N_{n_N}\rho {M^N_{n_N}}^{\dagger}\cdots\right]{M^1_{n_1}}^{\dagger}
\label{eq:krausMap}
\end{equation}
 In order to speed up the computation time, we use the 
procedure from Section \ref{section:OCT} to find the smallest set of Kraus operators from the Choi matrix. 

The pulse finder's fit function measures the overlap of the superoperator generated by the OCT pulse with the desired unitary, represented
by the gate fidelity \cite{fortunato02, nielsen},
\begin{equation}
F = \sum_k Tr(U^{\dagger}M_k)^2/2^d
\label{eq:pulseFidelity}
\end{equation}
where $U$ is the unitary we want to perform and $\{M_k\}$ is the non-unique set of Kraus operators that describe
the pulse superoperator \cite{pravia03}.  



We found two sets of OCT pulses: 
one including the electron $T_1$ and the cross relaxation processes ($T_x$), and one considering
only unitary evolution. 
The pulse found with optimal control but without including relaxation consists of two pulses ($p_n$) with a delay ($\tau_n$) between them:
\begin{equation}
(\tau_1) - \left(p_1\right) - (\tau_2) - \left(p_2\right) - (\tau_3)
\label{eq:hardOCTpulse}
\end{equation}
This composite pulse, which was found neglecting relaxation, has a fidelity with the desired unitary of 0.56.  

The OCT pulse found by including relaxation processes is three square pulses with delays as follows:
\begin{equation}
(\tau_1) - \left(p_1\right) - (\tau_2) - \left(p_2\right) - (\tau_3) - \left(p_4\right) - (\tau_4)	
\label{eq:hardOCTpulse}
\end{equation}
The computed fidelity for this pulse is 0.74.  It appears that the additional square pulses in the OCT sequences correct errors that hard pulses do not. 
These fidelities may appear low, but it should be noted that they correspond to a single application of the composite pulse.  We need only to saturate a transition in order to transfer polarization.  Therefore the angle of the pulse does not matter as it is repeated many times in the saturation train, and we can still achieve large polarization enhancements.  Additionally, these fidelities are calculated for the unitary that acts on the full electron-nuclear space.  For DNP we are not concerned with the full map that acts on the electron-nuclear system, but we are interested only in the action of the control pulses on the nuclear subspace.
Over the length of the saturation train the reduced map on the nuclear spin after tracing out the electron is essentially a polarizing channel for all sets of the 
pulses found.  This map takes the identity state to a polarized state as follows,
\begin{equation}
\Lambda(\mathds{1}) = \mathds{1} +pZ
\label{eq:polarizingChannel}
\end{equation}
with  the value of $p$ determining the final nuclear polarization for each type of 
pulse.  The OCT pulses found including the full open system dynamics act as the strongest polarizing channel and have a fidelity of 0.95 with the reduced map produced
by the ideal Overhauser effect. In contrast the fidelity of the reduced system dynamics with the ideal case is 0.66 for the closed system OCT pulses
and 0.31 for hard pulses.

\begin{figure*}
		\caption{Supermatrices for OE (a) and SE (b) as well as for the three sets of pulses: open OCT (c), closed OCT (d), and hard pulses (e). Dark Red indicates elements with magnitude 1, and at the other side of the spectrum dark blue elements are equal to 0. The operator describing the open OCT pulses has the best overlap with that for the OE.  The two other pulses have large contributions from components not present in (a), indicating that they are less effective at isolating the OE}	
				\label{fig:superoperators2}
	\[\arraycolsep=0.4pt\def\arraystretch{-1}
		\begin{array}{clcl}
		(a)&&(b)\\
		 \!(c)\!&\includegraphics[width=.5\linewidth]{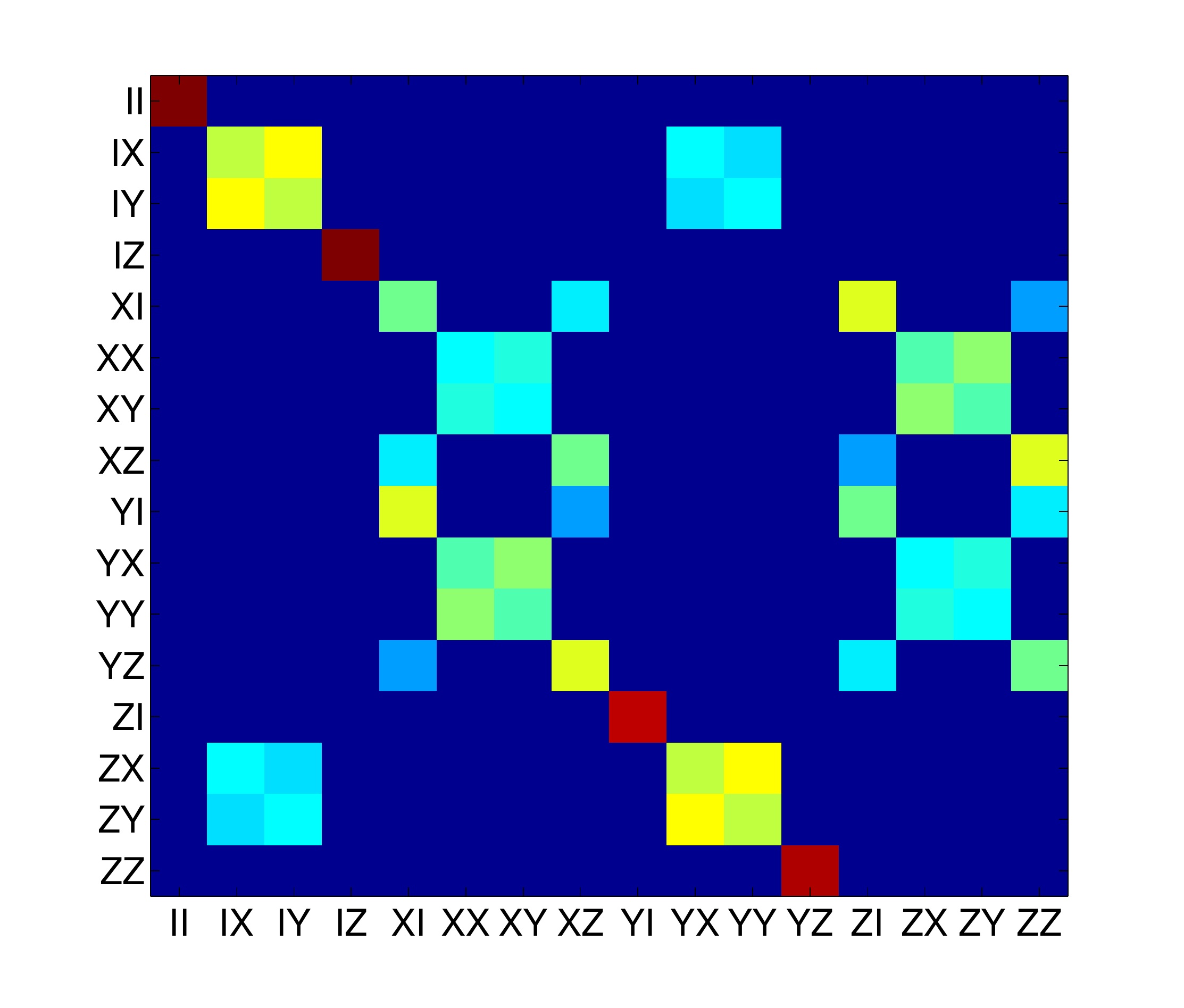} &\!(d)\!&\includegraphics[width=.5\linewidth]{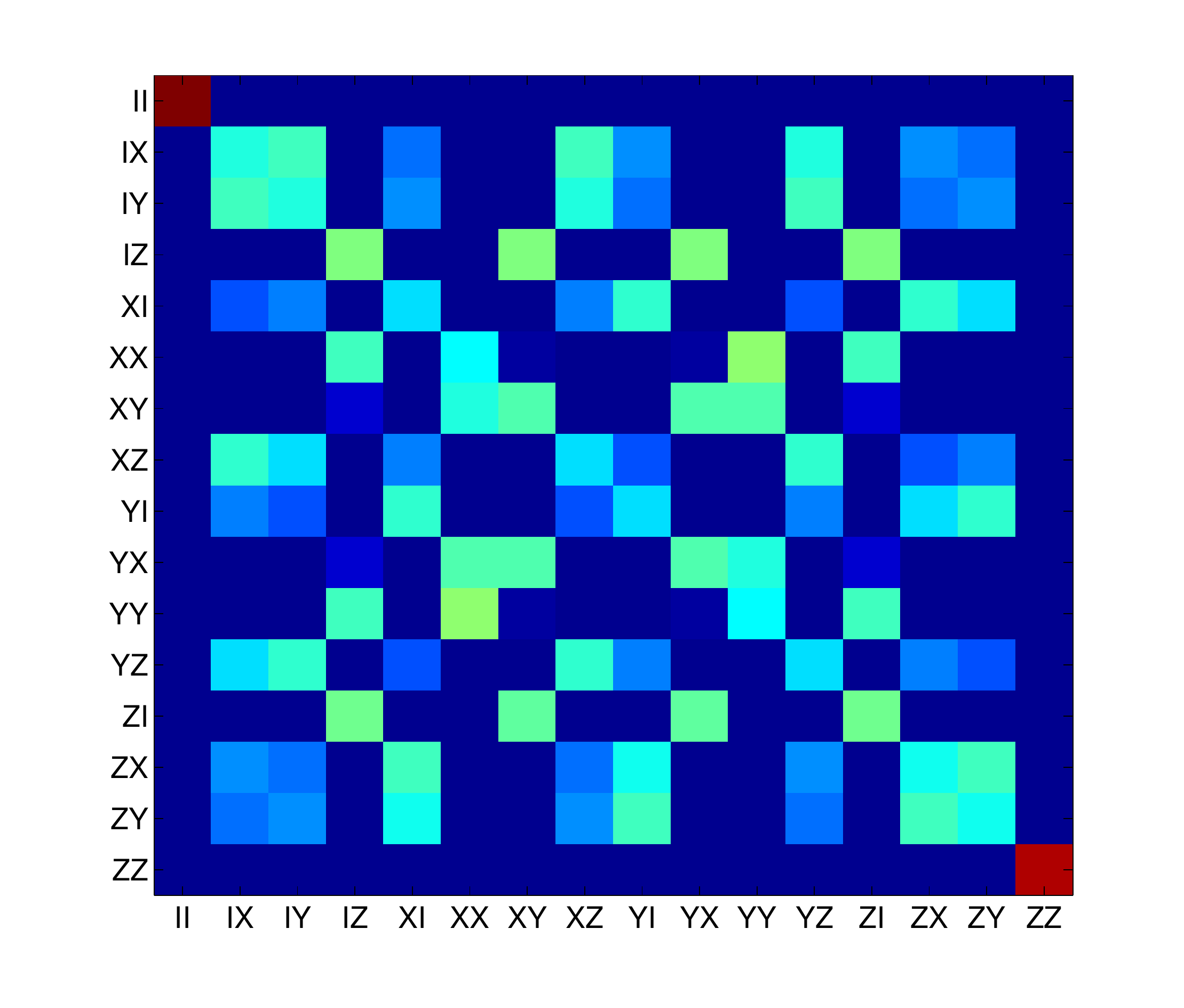}\\
	\!(e)\!&\includegraphics[width=.5\linewidth]{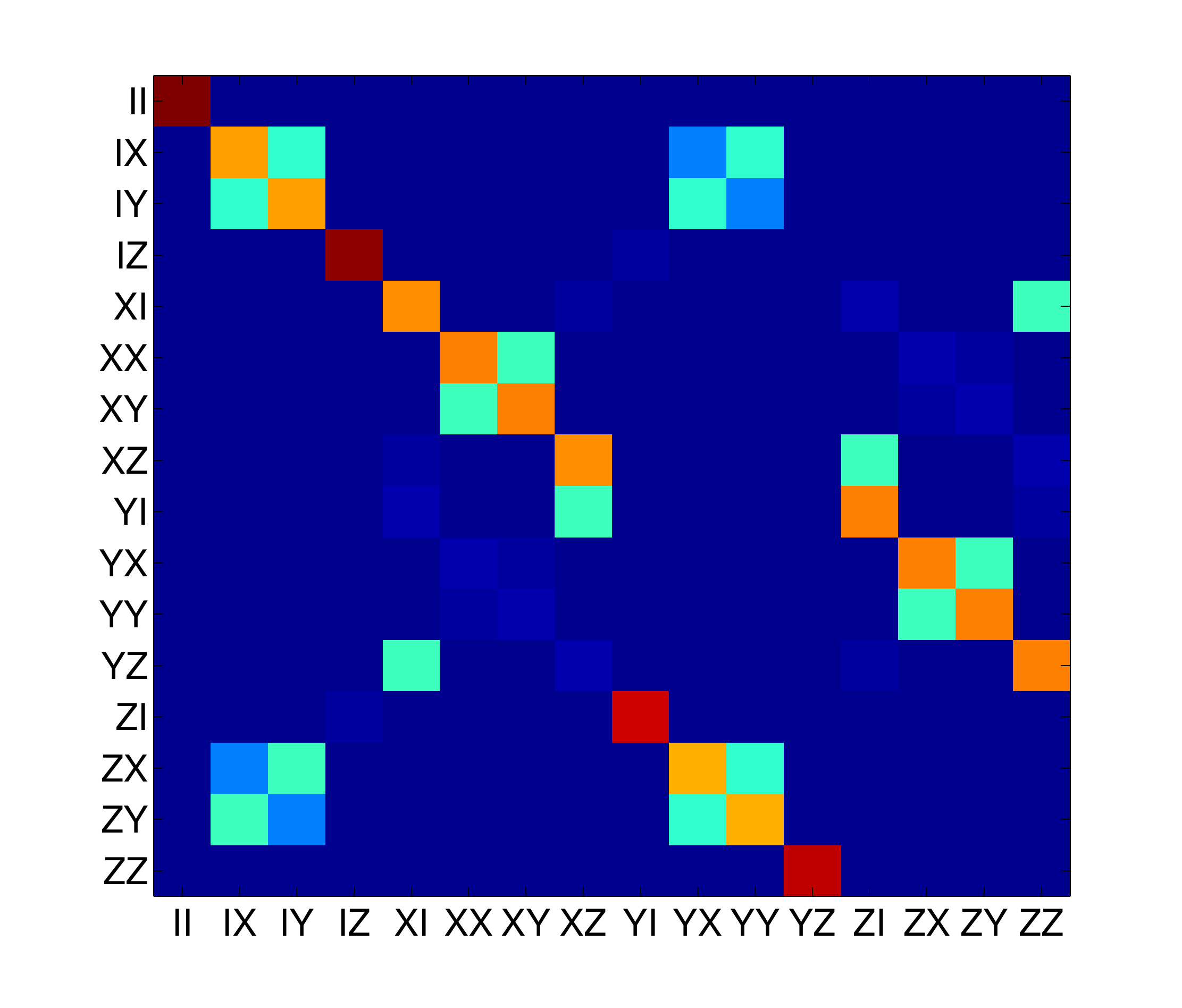} &&\includegraphics[width=.5\linewidth]{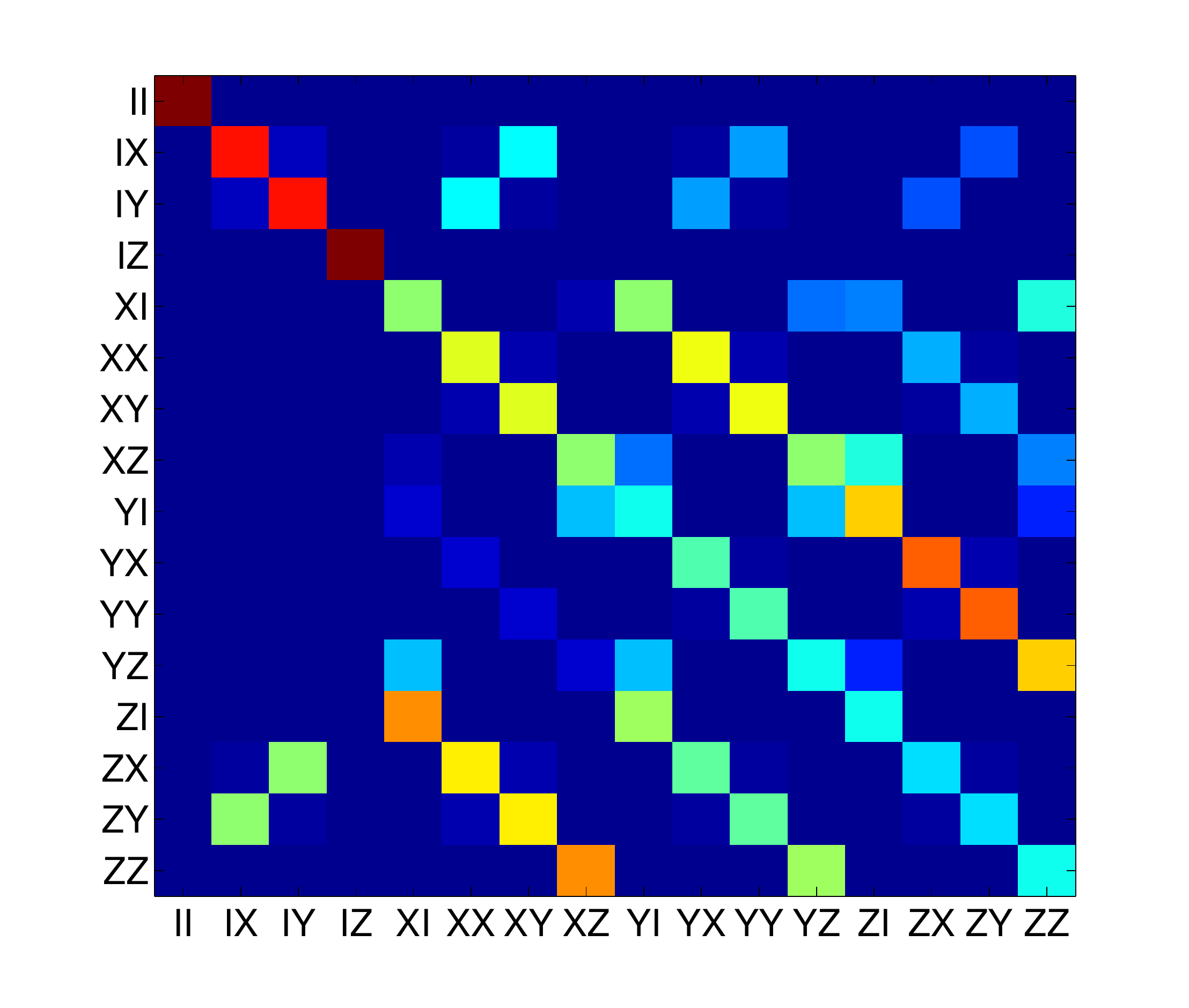}\\
		&\hspace{0.25cm}\includegraphics[width=.5\linewidth]{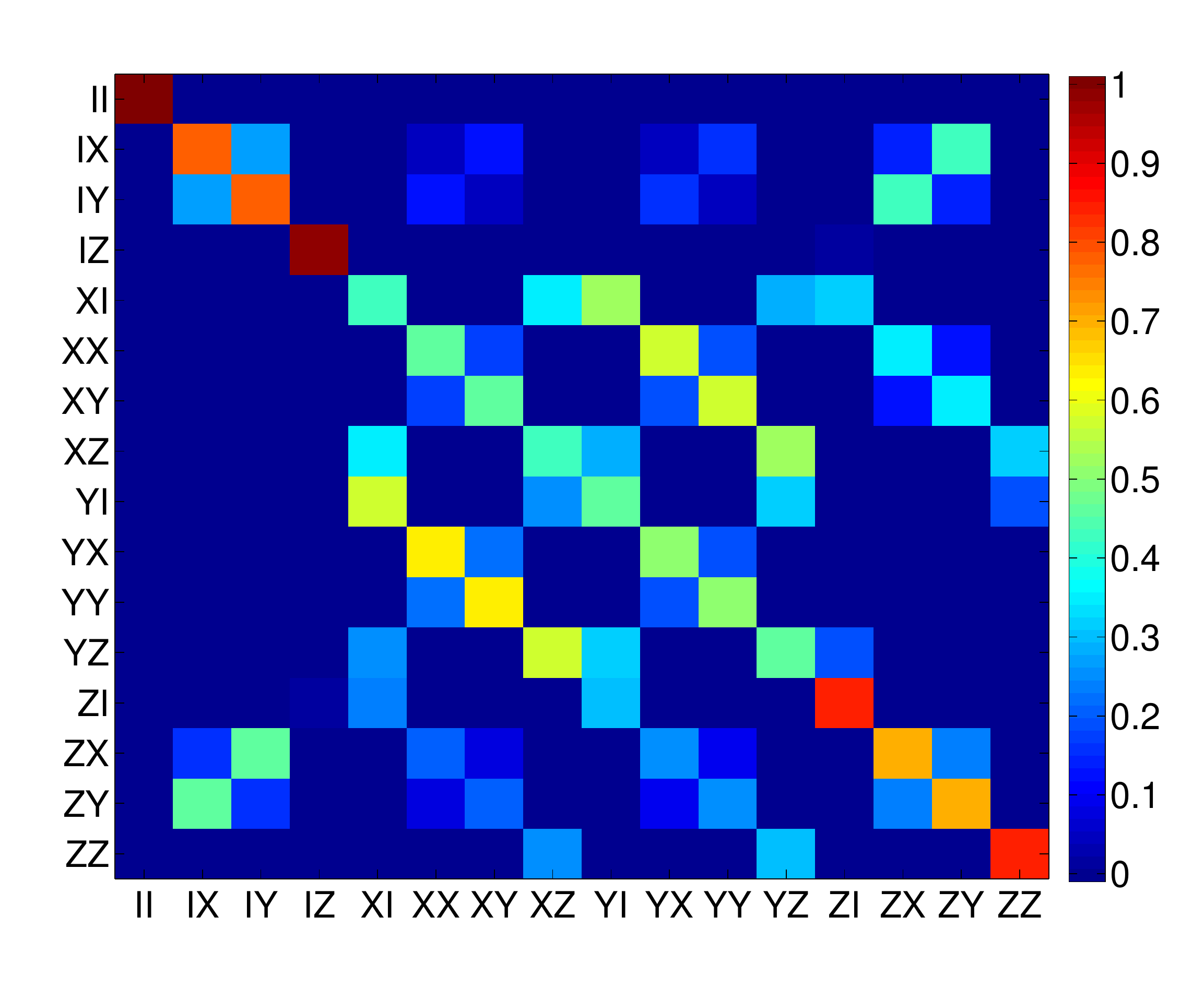}&&
		\end{array}
		\]

\end{figure*}

To analyze the map on the full space for each of the pulses, we rewrite it in the supermatrix formalism. 
Fig. \ref{fig:superoperators2} gives the superoperator matrices, $\textbf{S}$, in the
Pauli basis for the solid effect, Overhauser effect, both sets of OCT pulses, and hard pulses. The eigenvector of $\textbf{S}$ with eigenvector $1$ is equal to the final state into which the superoperator drives the system.  These
final states for the Overhauser effect and solid effect superoperators are given in Table \ref{table:finalStates}.
The $IZ$ and $ZI$ terms are proportional to the final nuclear and electron polarizations, respectively.  

The final state coefficient indicate that the OE drives the system to a positive nuclear polarization while depolarizing the electron spin.  In contrast, the SE results in a negatively polarized nuclear state but does not perform a strong depolarizing channel on the electron spin.  Additionally the SE 
creates coherences in the zero quantum terms, XY and YX, while the OE only leads to single quantum coherences in the 
final density matrix.

\begin{table}
	\centering 
		\begin{tabular}{l|c|c|c|c|c}\hline\hline
		& OE & SE & Open OCT & Closed OCT & Hard Pulses\\ \hline
		II& 0.25 & 0.25 & 0.25 & 0.25 & 0.25\\ 
		IX& 0 & 0 & 8.13e-8 & -9.25e-8 & 5.46e-8\\
		IY& 0 & 0 & -1.43e-7 & -1.76e-7 & 5.53e-11\\
		IZ& 1.88e-4 &  -1.85e-4 & 1.79e-4 & 1.24e-4 & 5.90e-5\\
		XI& 0 & 0 & -1.69e-5 & -3.98e-5 & 3.88e-5\\ 
		XX& 0 & 0 & -6.29e-12 & 1.38e-11 & 8.80e-12\\
		XY& 0 & -1.28e-6 & 9.93e-12 & 2.94e-11 & 1.36e-14\\
		XZ& 0 & 0 & -1.25e-8 & -2.06e-8 & 9.61e-9\\
		YI& 5.13e-6 & 0 & 2.64e-6 & 2.68e-6 & -5.70e-6\\
		YX& 0 & 1.28e-6 & 5.01e-13 & -8.52e-13 & -1.07e-12\\
		YY& 0 & 0 & -1.02e-12 & -2.04e-12 & -3.43e-14\\
		YZ& 3.85e-9 & 0 & 2.07e-9 & 1.44e-9 & -1.27e-9\\
		ZI& -6.52e-6 & -1.88e-4 & -1.35e-5 & -6.66e-5 & -1.30e-4\\
		ZX& 0 & 0 & -4.29e-12 & 2.51e-11 & -2.94e-11\\
		ZY& 0 & 0 & 9.01e-12 & 4.92e-11 & 6.99e-15\\
		ZZ&	 1.20e-5 &  1.40e-7 & -9.92e-9 & -3.42e-8 & -3.16e-8\\
		\hline\hline
		\end{tabular}
		\caption{Final states given by the superoperators for the Overhauser effect and solid effect compared to the states driven by the superoperators for the three sets of pulses found.}	
		\label{table:finalStates}
\end{table}

From the superoperator matrices in Fig. \ref{fig:superoperators2} we can see qualitatively that the superoperator for OCT pulses
has the same structure as the superoperator for the Overhauser effect with perfect $\pi/2$ rotations on the electron only transitions.
The closed system OCT pulses and hard pulses are also similar in structure to the OE map, but have increasing contributions from terms not present in the OE map.

\section{simulations}\label{section:simulations}



\begin{figure}[!bt]
	\centering
		\includegraphics[width=8cm]{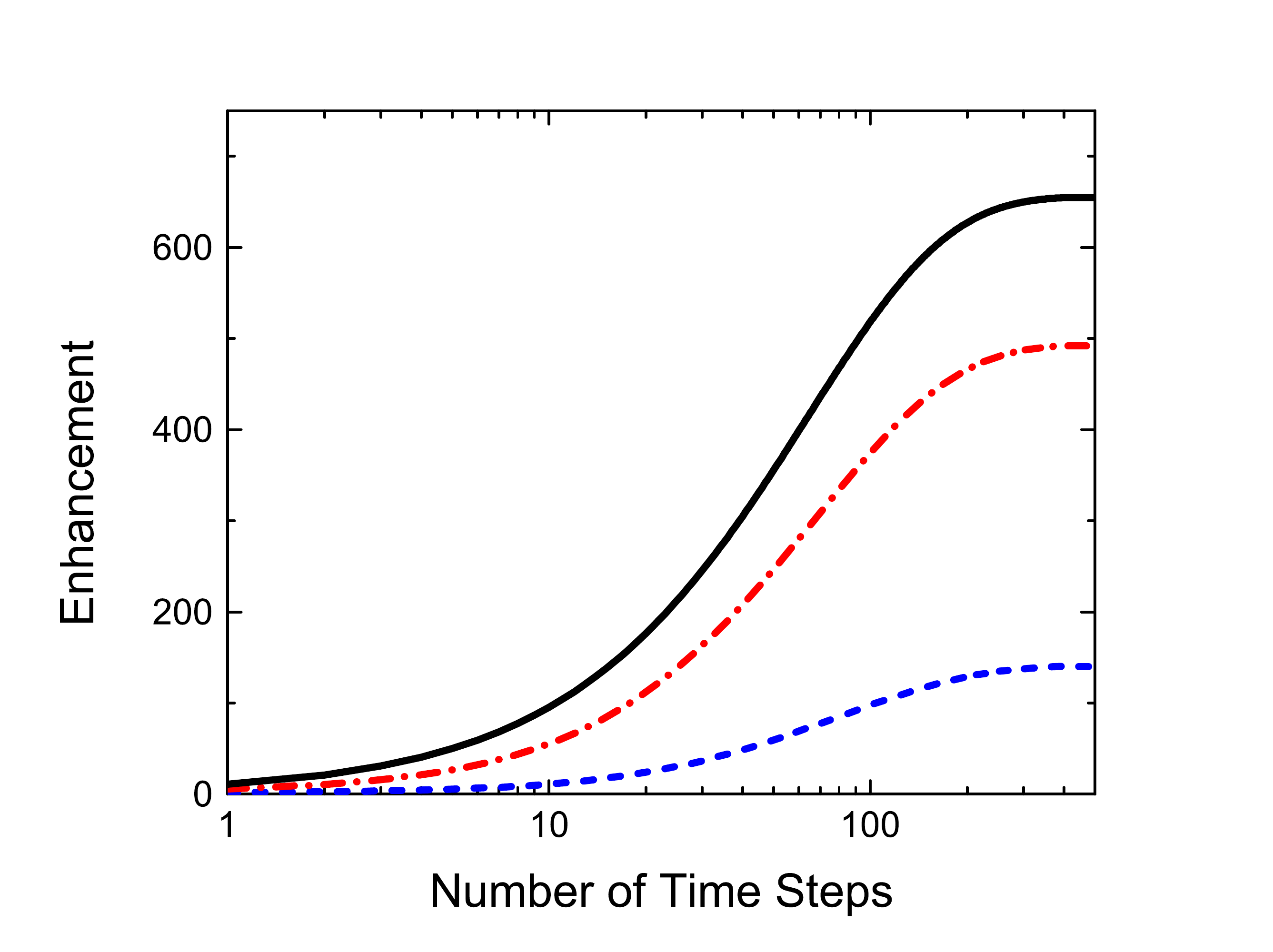}
	\caption{Comparison of OCT pulses to hard pulses.  Black solid line is saturation by open system OCT pulses, dashed red is
	closed system OCT pulses, and short dashed blue is saturation with hard pulses.}
	\label{fig:8MHzPulseComparisonSimulation}
\end{figure}

Figure \ref{fig:8MHzPulseComparisonSimulation} shows the results of simulating the OCT and non-OCT pulses.  
The simulations show that both sets of OCT pulses (solid black for open system OCT and dot-dashed red for closed system OCT) give a greater enhancement than
 hard $\pi/2$ pulses (dashed blue).  The open system OCT pulses yield 
a nuclear polarization of 654, almost the full enhancement. The closed system OCT pulses produce a nuclear polarization
491, while the enhancement due to hard pulses is 140.   

\begin{figure}[h]
	\centering
	$\begin{array}{cc}
           (a)&\\
					&\hspace{-.5cm}\includegraphics[width=0.4\textwidth]{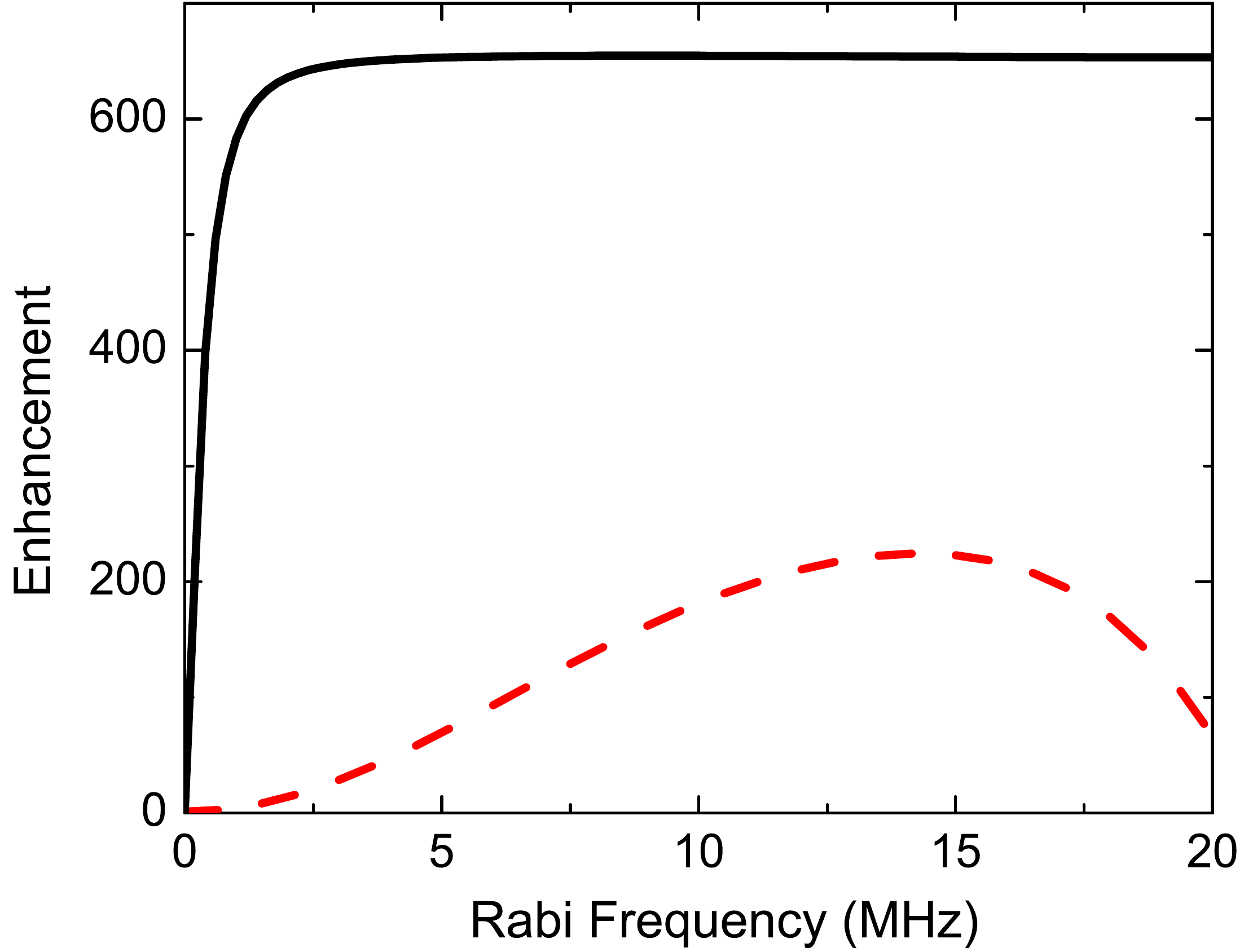}\\
					(b)&\\
					&\hspace{0cm}\includegraphics[width=0.43\textwidth]{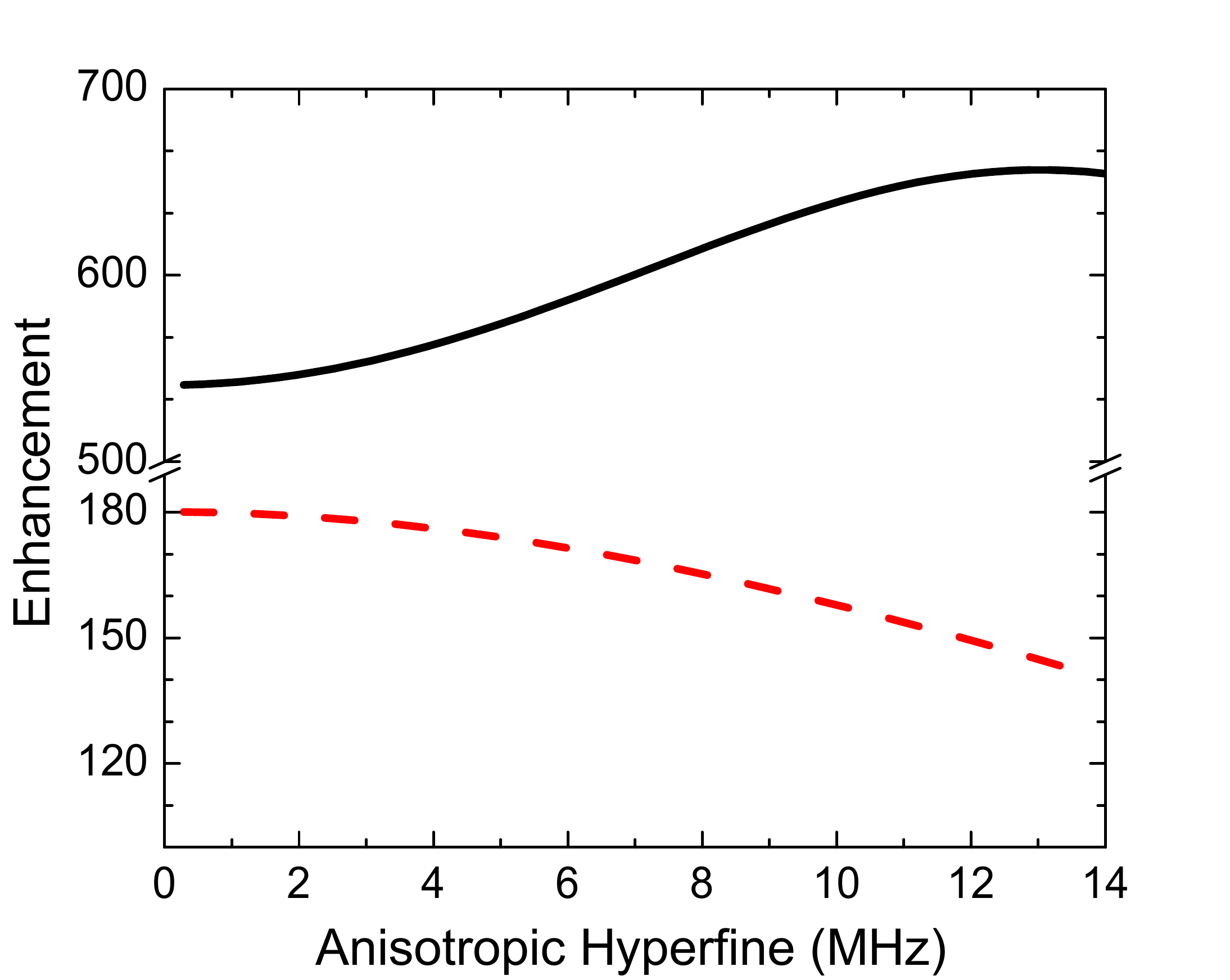}\\
  \end{array}$
	\caption{Simulations of pulses varying the Rabi frequency (a) and the anisotropic hyperfine coupling (b). The solid black line indicates simulations of the OCT pulses, red dashed is hard pulses.}
	\label{fig:pulseVariations}
\end{figure}

We simulated the performance of the open system OCT pulse  with variations in both Rabi frequency and the anisotropic hyperfine coupling and compared
it to hard pulses with the same variations
(see Fig. \ref{fig:pulseVariations}).  The OCT pulse was optimized for a Rabi frequency of 8 MHz, the frequency experimentally achievable 
at a reasonable power.  The simulations show that the OCT pulses are quite robust to changes in Rabi frequency, while hard pulses reach the maximum polarization
enhancement at a Rabi frequency equal to anisotropic hyperfine coupling (14 MHz in this simulation).  The enhancement falls of relatively quickly for hard pulses
as compared to the OCT pulses. 

Fig. \ref{fig:pulseVariations}(b) presents  the results of varying the anisotropic hyperfine frequency, $B$ (as in the Hamiltonian of Eq. \ref{eq:driftHamiltonian}). The OCT pulses give the highest enhancements when the hyperfine
coupling is near maximal, where there is the largest amount of nuclear state mixing.  The hard pulses actually improve with lower anisotropic hyperfine values. 
This result is somewhat nonphysical, however, as the cross relaxation rate is the same for all values of $B$.  In reality the cross relaxation is weaker when
$B$ is smaller.  In these simulations where $T_x$ is constant and $B$ decreases, the hard pulses improve because the solid effect becomes less likely as the 
zero quantum transitions is more strongly forbidden, while the Overhauser effect is still allowed through the constant cross relaxation. 

\begin{figure}
	\centering
		\includegraphics[width=8cm]{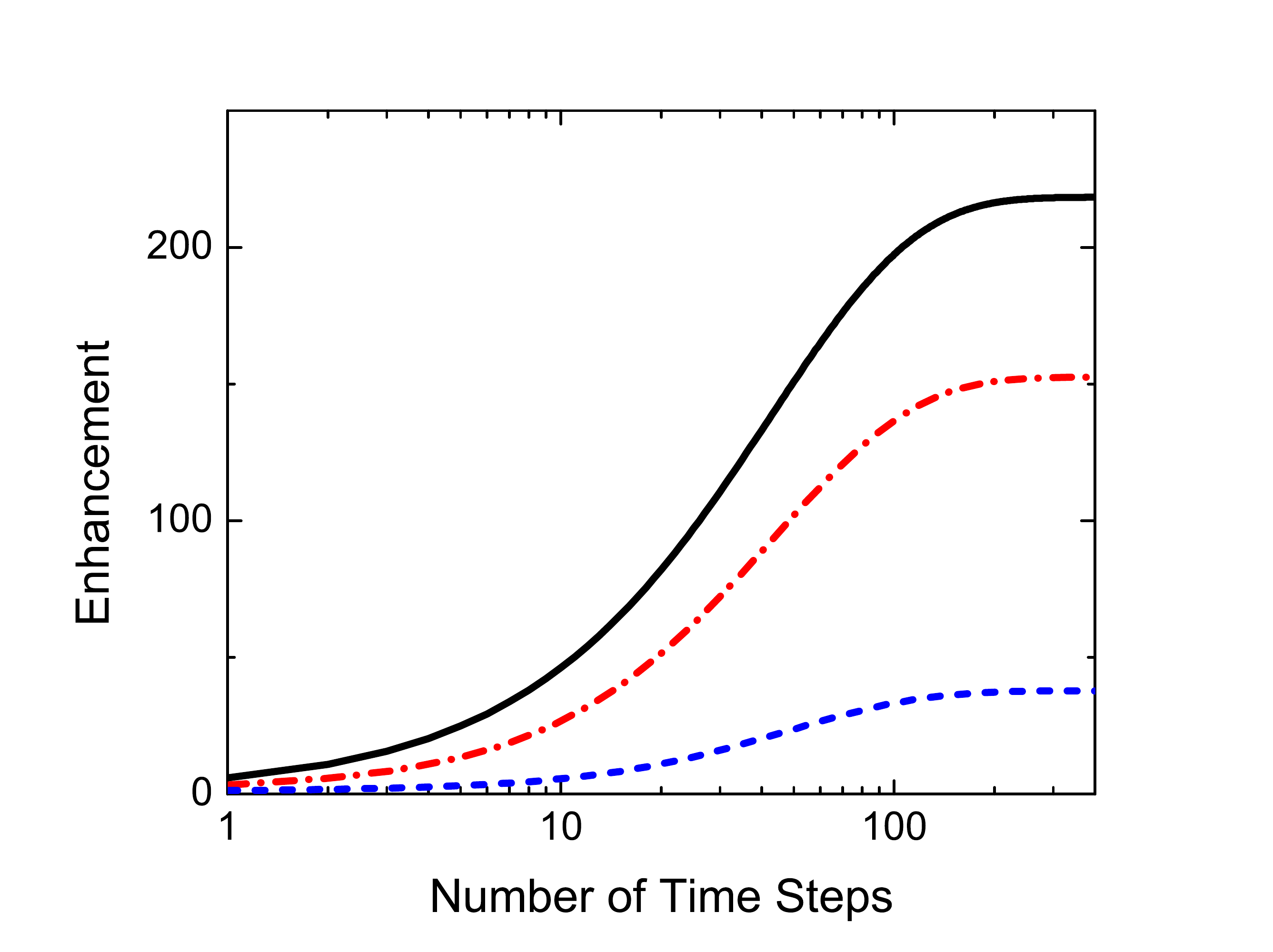}
	\caption{DNP with double quantum relaxation.  The black solid line is open OCT pulses, red dashed closed OCT, and blue dotted is hard pulses.  Here $T_{dq} = 2T_{zq}.$}
	\label{fig:doubleQuantumComparison}
\end{figure}

One possible leakage pathway in the physical system is the double quantum transition.  While relaxation through flip-flip
interactions will be a weaker process than the flip-flop that drives DNP, both pathways are present.  We have simulated DNP in a system
with a double quantum relaxation rate of 0.5 times the zero quantum rate.  The results are shown in Fig \ref{fig:doubleQuantumComparison}.
When double quantum relaxation is included, the DNP enhancements are lowered for all pulses: 218 for open OCT, 152 for close OCT, and 38 for hard pulses.

It should be noted
that there is a significant difference between the simulated system and the physical system.  In the simulations we are looking
only at the two spin system consisting of one electron and one proton with the couplings found in malonic acid.  We find
the DNP enhancement of the coupled proton.  In the experiment, however, we observe the bulk proton signal, as the number of 
protons coupled directly to an electron is too small to detect directly.  The actual experimental nuclear polarization depends on the couplings between the bulk protons and the hyperfine
coupled protons, and the build up times are limited by spin diffusion \cite{ramanathan08}. 



\section{experiment}

The system used in this work is irradiated malonic acid, an organic radical with one electron hyperfine coupled to one proton. Malonic acid
is a well characterized sample widely used in ESR \cite{dalton71,sagstuen00}. 
The malonic acid sample used was a single crystal irradiated for 5 hours with 8 keV X-rays and subsequently annealed at $45^{\circ}$C for
15 hours. 

The Hamiltonian for this two spin system is the drift Hamiltonian of Eq. (\ref {eq:driftHamiltonian}).
In an external field of 3406 G, the electron Larmor frequency is 9.59 GHz, the nuclear resonance is
14.57 MHz \cite{mccalley,sagstuen00}.  The terms $A$ and $B$ depend on the orientation of the crystal in the external magnetic field.  
In the orientation that maximizes the mixing of the nuclear states, the isotropic ($A$) and anisotropic ($B$) parts 
of the hyperfine coupling are -42.7 MHz and 14.7 MHz respectively.  We determined 
this orientation in a CW-ESR spectrometer.

\begin{figure}
	\centering
		\includegraphics[scale=0.22]{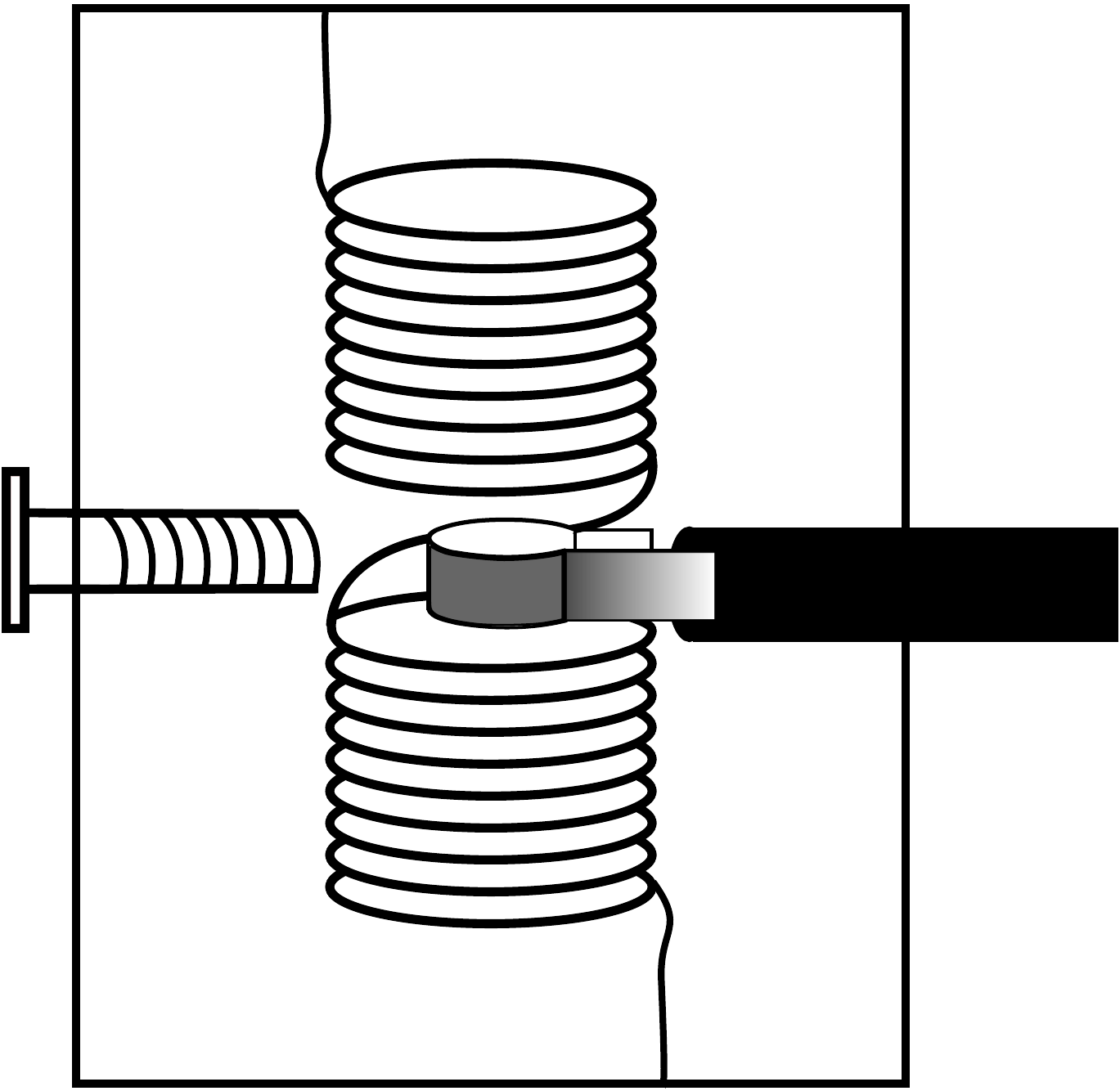}
		\hspace{10mm}
		\includegraphics[scale=0.22]{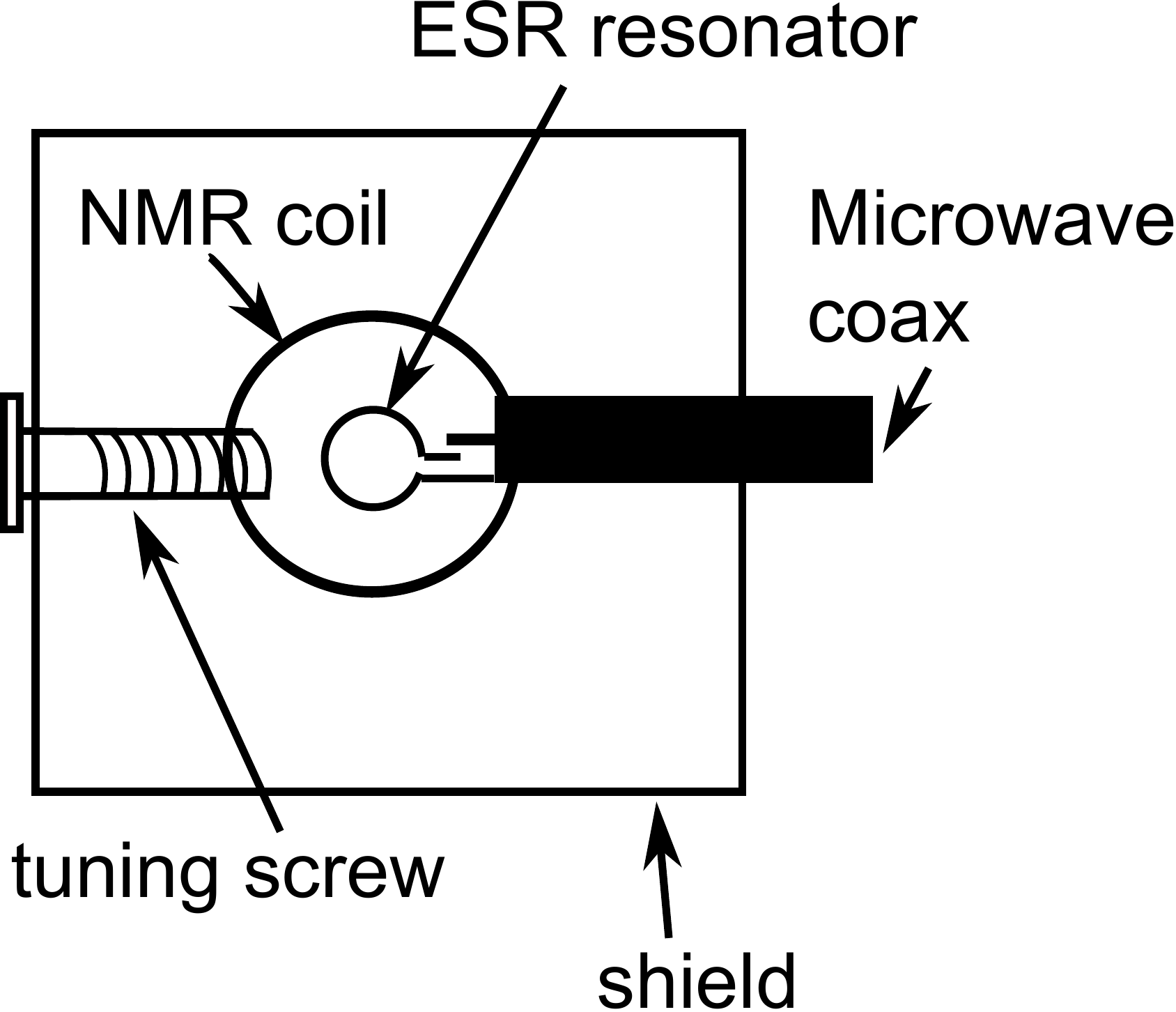}
	\caption{Probe Diagram: the figure on the left shows the side view of the probe, and on the right is the top view.  The ESR resonator in the shape of a loop-gap
	is concentric to the NMR coil.}
	\label{fig:probeDiagram}
\end{figure}

We built a double frequency probehead with an ESR resonator surrounded by a split coil for NMR detection.  
The schematic for the probe is shown in Fig \ref{fig:probeDiagram}.  The ESR resonator is based on the standard loop-gap resonator design, but is 
easier to construct than typical bridged loop-gap resonators commonly used in DNP experiments \cite{shane98,forrer90}.

The resonator is soldered onto the outer conductor of a microwave coax at one end, while the 
other end is open (see top view of Fig \ref{fig:probeDiagram}).  A tuning screw on the opposite side of the shield allows tuning and matching of the ESR resonance
over a 1 GHz range around 10 GHz.  


We performed three sets of experiments, all using the basic pulse sequence given in Figure 	~\ref{fig:OCTDNPexperiment}, with hard pulses, closed system OCT,
and open system OCT pulses as the $\pi/2$ saturation pulse.  
To measure the thermal signal without DNP, we used an Ernst angle detection with 400,000 scans.  Due to the large dipolar coupling between protons
in malonic acid, $T_2$ of the proton signal was less than the deadtime of the probe.  The NMR signal was detected with a magic echo sequence to overcome this issue. 
The proton linewidth was 40 kHz.

\section{results}

\begin{figure}[!bt]
	\centering
		\includegraphics[width = 8cm]{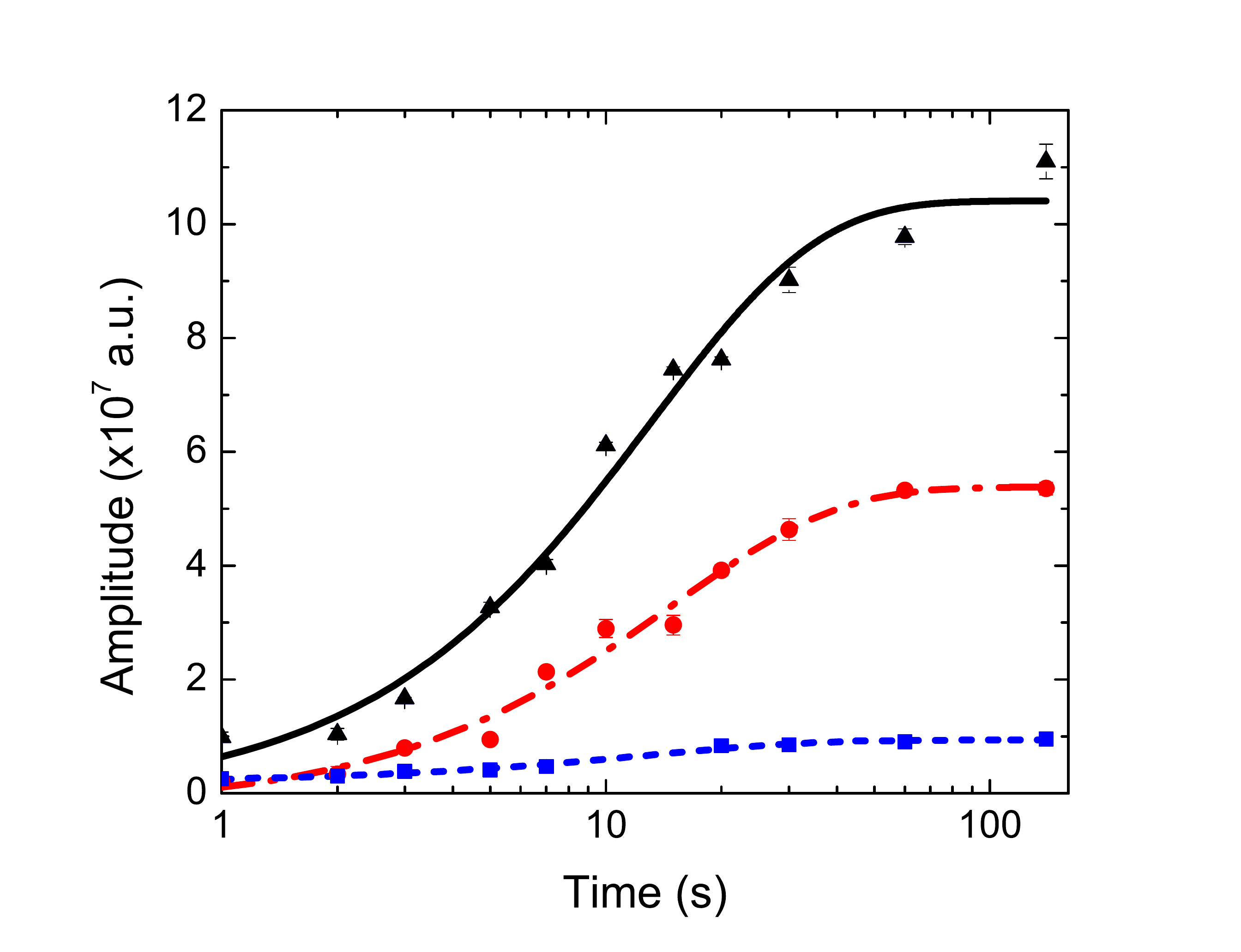}
	\caption{Results of the experiment with fits of the buildup curves.  Black triangles are the data from saturation with open system OCT pulses, red circles
	are for closed OCT pulses, and blue squares are for hard pulses.}
	\label{fig:OpenClosedOCTComparison}
\end{figure}

Figure \ref{fig:OpenClosedOCTComparison} shows the build up curves for the three sets of pulses.  
The open system OCT pulses produce the highest nuclear polarization, 11.1 times the enhancement produced by hard pulses,
and 2.2 times that produced by the OCT pulses found without relaxation (closed system pulses).  The closed system pulses yield a polarization 5.1 times that
of hard pulses. These ratios are consistent with those found in the simulations.  
The maximum polarization enhancement achieved with the open OCT pulses was a factor of $180\pm 36$. 


From fits of the data in Figure \ref{fig:OpenClosedOCTComparison} we find the build up times for each curve.  For all three curves the rate determining 
process is spin diffusion, as we detect the bulk proton signal and do not measure the polarization of the spins directly coupled to the electron defects. 
The hard $\pi/2$ pulses
produce a build up time of $12.7 \pm 1.8$s , the closed OCT build up is $14.9 \pm 1.7$s, and $13.2\pm 1.5$s for the open OCT pulses.  

\begin{figure}[h]
	\centering
		\includegraphics[width = 8cm]{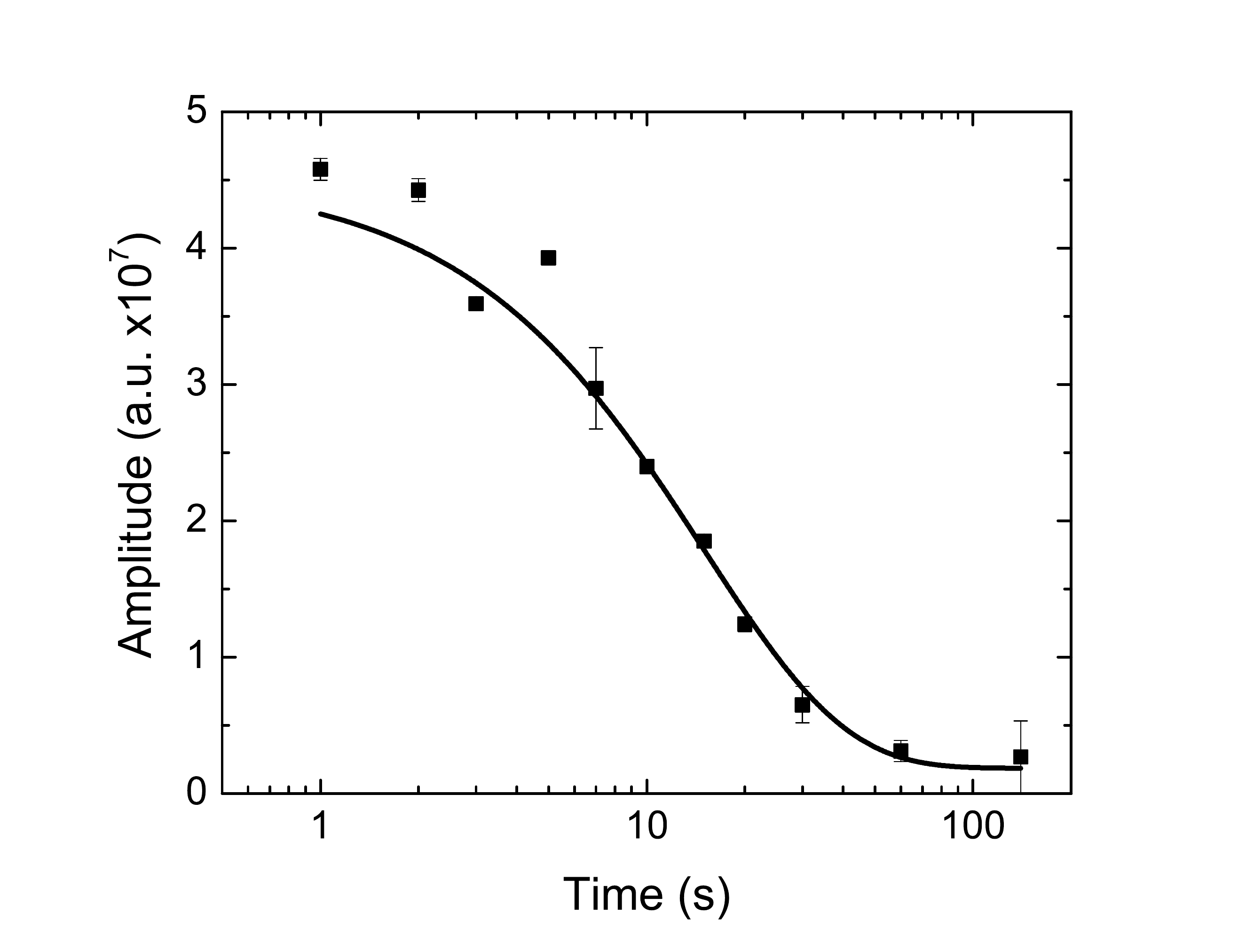}
	\caption{Effective $T_1$ measurement after polarization for 10s. Fit to a single exponential given by black line.  $T_1$ measured to be $10.5\pm1.6$s}
	\label{fig:EffectiveT1measurement}
\end{figure}

Figure \ref{fig:EffectiveT1measurement} shows the measured effective $T_1$ of the bulk nuclear spins after DNP.  A fit of this 
curve to a single exponential decay gives $T_1 = 10.5\pm1.6$ s.  Like the DNP process, the return to thermal equilibrium depends on spin diffusion.  The fast electron $T_1$ process will depolarize
nuclear spins in its vicinity, which will then diffuse through the bulk.  Thus we expect the effective $T_1$ of the bulk nuclear spins to
be on the same order as the polarization build up times.  
We attribute the slight difference between the polarization and depolarization times 
to the fact that there are more pathways for the bulk spins to relax through than there are pathways for polarizing.

These findings provide an example of how optimal control theory can improve polarization transfer in DNP experiments, and, more generally, provide superior
control in open quantum systems.  Even with a relatively simple pulse sequence with few time steps we can achieve an order of magnitude increase
in polarization using OCT pulses versus hard pulses.

While the OCT pulses do produce a nuclear polarization enhancement of 180, this increase is less than the theoretical maximum enhancement
of 660 and less than the simulated value.  
The inability to reach this level of enhancement could be due to several factors already discussed in Section \ref{section:simulations}, such as leakage pathways or 
asymmetric saturation of the electron resonances.  
The simulations  presented earlier have shown that leakage through double quantum relaxation (in Fig. \ref{fig:doubleQuantumComparison}), 
or  asymmetric saturation (\ref{fig:DNPspace2} )will also reduce the final nuclear polarization.  We have also seen
that the hard pulses are more susceptible to such deviations from the idealized case, and it is not surprising that the hard pulses are worse than the simulations
by a factor of 10, while OCT pulses are only a factor of 5 different than their simulated value.

\section{discussion}

These results could be extended to find control sequences for larger spin systems. In particular we have neglected any nuclei further from the electron
defect than the nearest proton.  These nuclear spins may still have small hyperfine interactions with the electron spin, and we have not allowed any 
transfer pathways in the pulse optimization which might polarize these spins directly.  We could also include the dipolar couplings between nuclear spins
to simulate spin diffusion, the rate limiting step in this experiment.  

It may also be useful to design control pulses for samples
other than single crystals.  In the case presented here the two spin system was assumed to be in the same orientation throughout the sample.  It may be 
particularly useful to consider DNP in powder samples.  This
would require optimizing pulses for a set of internal Hamiltonians.  

This method of pulse finding is useful for quantum control in systems where 
relaxation plays a strong role in the dynamics. The Kraus operator description is particularly convenient as one can write a modular pulse finding code 
in which decoherent processes can be added to the superoperator successively.
In principle another superoperator formalism, such as the Lindblad representation under the condition that the system is Markovian, should give the same results.  
Including relaxation processes in the pulse finder produces pulses that select the desired transitions 
more accurately than using a superoperator that accounts only for unitary processes. 
This experiment demonstrates that we can design control sequences for open quantum systems that account for interactions with the environment. 


Not only is it possible to create control operations in dissipative systems, but we have shown that an open quantum system approach can provide significant 
improvements over unitary control. We anticipate that superoperator optimized pulse design will lead to superior control in spin based systems and 
other quantum devices.

 \section*{Acknowledgments}
This work was supported by the Natural Sciences and Engineering Research
Council of Canada (NSERC), Canadian Excellence Research Chairs (CERC)
Program and the Canadian Institute for Advanced Research (CIFAR) and
Industry Canada.  

We acknowledge many useful discussions with Troy Borneman 
and Sekhar Ramanathan.

\bibliography{DNPpaper}

\end{document}